\documentclass{aa}  
\makeatletter
\renewcommand*\aa@pageof{, page \thepage{} of \pageref*{LastPage}}
\usepackage{graphicx}
\usepackage{booktabs}
\usepackage{placeins}

\usepackage{txfonts}

\usepackage[
    colorlinks=true,
    allcolors=blue
    ]{hyperref}

\usepackage{natbib,twoopt}
\usepackage[export]{adjustbox}
\usepackage{siunitx}

\defcitealias{mckean_etal:2016}{MK16}
\newcommand\mksixteent{\citetalias{mckean_etal:2016}}
\newcommand\mksixteenp{\citepalias{mckean_etal:2016}}

\begin{document} 
   \title{Spectral modelling of Cygnus\texorpdfstring{\,}{}A between 110 and 250\texorpdfstring{\,}{}MHz}

   \subtitle{Impact on the LOFAR 21-cm signal power spectrum}

   \author{E. Ceccotti
          \inst{1,2}\fnmsep\thanks{E-mail: emilio.ceccotti@inaf.it}
          \and A.~R.~Offringa\inst{3,1}
          \and L.~V.~E.~Koopmans\inst{1}
          \and F.~G.~Mertens\inst{4}
          \and M.~Mevius\inst{3,1}
          \and A.~Acharya\inst{5}
          \and S.~A.~Brackenhoff\inst{1}
          \and B.~Ciardi\inst{5}
          \and B.~K.~Gehlot\inst{1}
          \and R.~Ghara\inst{6}
          \and J.~K.~Chege\inst{1}
          \and S.~Ghosh\inst{1}
          \and C.~H\"{o}fer\inst{1}
          \and I.~Hothi\inst{4,7} \and \\
          I.~T.~Iliev\inst{8}
          \and J.~P.~McKean\inst{1,9,10}
          \and S.~Munshi\inst{1}
          \and S.~Zaroubi\inst{11,1,5}
          }

   \institute{
            Kapteyn Astronomical Institute, University of Groningen, PO Box 800, 9700 AV Groningen, The Netherlands 
         \and
            INAF -- Istituto di Radioastronomia, Via P.~Gobetti 101, 40129 Bologna, Italy
        \and
             ASTRON, PO Box 2, 7990 AA Dwingeloo, The Netherlands
        \and
            LERMA, Observatoire de Paris, PSL Research University, CNRS, Sorbonne Université, F-75014 Paris, France
        \and
            Max-Planck Institute for Astrophysics, Karl-Schwarzschild-Straße 1, 85748 Garching, Germany
        \and
            Department of Physical Sciences, Indian Institute of Science Education and Research Kolkata, Mohanpur, WB 741246, India
        \and
            Laboratoire de Physique de l'ENS, ENS, Universit\'{e} PSL, CNRS, Sorbonne Universit\'{e}, Universit\'{e}e Paris Cit\'{e}, 75005 Paris, France
        \and
            Astronomy Centre, Department of Physics \& Astronomy, Pevensey III Building, University of Sussex, Falmer, Brighton, BN1 9QH, United Kingdom
        \and
            South African Radio Astronomy Observatory (SARAO), PO Box 443, Krugersdorp 1740, South Africa
        \and
            Department of Physics, University of Pretoria, Lynnwood Road, Hatfield, Pretoria, 0083, South Africa
        \and
            Department of Natural Sciences, The Open University of Israel, 1 University Road, PO Box 808, Ra’anana 4353701, Israel
             }
   \date{Received 21 November 2024; accepted 24 February 2025}

  \abstract{
  Studying the redshifted 21-cm signal from the the neutral hydrogen during the Epoch of Reionization and Cosmic Dawn is fundamental for understanding the physics of the early universe. One of the challenges that 21-cm experiments face is the contamination by bright foreground sources, such as Cygnus\,A, for which accurate spatial and spectral models are needed to minimise the residual contamination after their removal. In this work, we develop a new, high-resolution model of Cygnus\,A using Low Frequency Array (LOFAR) observations in the 110--250\,MHz range, improving upon previous models by incorporating physical spectral information through the forced-spectrum method during multi-frequency deconvolution. This approach addresses the limitations of earlier models by providing a more accurate representation of the complex structure and spectral behaviour of Cygnus\,A, including the spectral turnover in its brightest hotspots. The impact of this new model on the LOFAR 21-cm signal power spectrum is assessed by comparing it with both simulated and observed North Celestial Pole data sets. Significant improvements are observed in the cylindrical power spectrum along the Cygnus\,A direction, highlighting the importance of having spectrally accurate models of the brightest foreground sources. However, this improvement is washed out in the spherical power spectrum, where we measure differences of a few hundred mK at $k<0.63\,h\,\text{cMpc}^{-1}$, but not statistically significant. The results suggest that other systematic effects must be mitigated before a substantial impact on 21-cm power spectrum can be achieved.}


   \maketitle
%

\section{Introduction}

If we look at the sky at 150\,MHz, we expect to find more than one source per $\text{deg}^2$ with a flux density lower than 1\,Jy \citep{wilman_etal:2008, intema_etal:2017}. Radio sources become increasingly rare as their brightness increases \citep[e.g.][]{franzen_etal:2019}, with only a few having flux densities higher than 100\,Jy. The brightest radio sources in constellations are named after their constellation with an `A' suffix, forming the so-called `A-team'. The most powerful A-team source at 150\,MHz is currently Cygnus\,A \citep[Cyg\,A; 3C\,405, RA $19^\text{h}59^\text{m}28\rlap{.}^\text{s}36$, Dec $+40^\circ44'02\rlap{.}''10$ in J2000;][]{baumgartner_etal:2013,charlot_etal:2020,vinyaikin:2014}, an FR-II radio galaxy with a flux density of about 11\,kJy at 150\,MHz, that, together with Cassiopeia\,A (Cas\,A), Taurus\,A (Tau\,A), and Virgo\,A (Vir\,A), forms the northern A-team \citep[e.g.][]{degasperin_etal:2020}.

The high brightness of the A-team sources is a double-edged sword for radio interferometric observations. On the one hand, these sources are so easy to measure even for low sensitivity instruments that their fluxes are well known, making them good (amplitude) calibrators, especially when unresolved. On the other hand, they can affect observations of a given target field even when they are tens of degrees away. This poses a challenge for wide-field instruments, such as LOFAR\footnote{Low-Frequency Array, \url{https://www.astron.nl/telescopes/lofar}} \citep{vanhaarlem_etal:2013} and MWA\footnote{Murchison Widefield Array, \url{http://www.mwatelescope.org}} \citep{lonsdale_etal:2009}, which have strong primary beam sidelobes. Different techniques have been developed to reduce such a contribution in the targetted observations. Examples are fast flagging of the data where the predicted contamination is above a certain threshold \citep{shimwell_etal:LoTSS_I:2017}, more computational demanding direction-dependent (DD) calibration \citep{gan_etal:2023}, demixing \citep{vandertol:2009} when the A-team source is more than $30^\circ$ away, or peeling \citep{noordam:2004,mitchell_etal:2008} when it is closer. In all of these cases, good spatial and spectral models of the A-team sources are required to minimise the residual contamination after the removal. 

Minimising the A-team contamination is particularly important for those experiments that aim to probe the neutral hydrogen (HI) during the Epoch of reionization (EoR) and Cosmic Dawn (CD) via the redshifted 21-cm line \citep[see e.g.][for some reviews on 21-cm cosmology]{furlanetto_etal:2006,pritchard_loeb:2012,liu_shaw:2020}. Because the current generation of instruments does not have the sensitivity to make direct images of the 21-cm signal, they mainly focus on estimating the spatial fluctuations of the 21-cm signal by measuring its power spectrum. Increasingly more stringent upper limits have been set on the 21-cm power spectrum from the EoR by radio interferometers such as LOFAR \citep{mertens_etal:2020} and MWA \citep{trott_etal:2020,kolopanis_etal:2023} with the highest frequency band, GMRT \citep{paciga_etal:2013}, PAPER\footnote{Precision Array for Probing the Epoch of Reionization, \url{http://eor.berkeley.edu}} \citep{parsons_etal:2010,kolopanis_etal:2019}, and HERA\footnote{Hydrogen Epoch of Reionization Array, \url{https://reionization.org/}} \citep{deboer_etal:2017,abdurashidova_et_al:2022}, and from the CD by instruments such as LOFAR \citep{gehlot_etal:2019} and MWA \citep{yoshiura_etal:2021} with the lowest frequency band, OVRO-LWA\footnote{Owens Valley Radio Observatory Long Wavelength Array, \url{https://leo.phys.unm.edu/~lwa}} \citep{greenhill_bernardi:2012,eastwood_etal:2019}, NenuFAR\footnote{New Extension in Nançay Upgrading LOFAR, \url{https://nenufar.obs-nancay.fr}} \citep{zarka_etal:2012,munshi_etal:2024:ul}, and AARTFAAC\footnote{LOFAR Amsterdam-ASTRON Radio Transients Facility and Analysis Centre, \url{http://aartfaac.org}} \citep{gehlot_etal:2020}. The next generation of instruments, such as HERA and SKA-Low\footnote{Square Kilometre Array, \url{https://www.skao.int/en/explore/telescopes}} \citep{dewdney_etal:2009,koopmans_etal:2015} are expected to have the sensitivity to detect the cosmological signal via power spectra, and SKA-Low has the potential to directly make tomographic images \citep[e.g.][]{giri_etal:2018a,giri_etal:2018b,bianco_etal:2024}. 

In all of the aforementioned cases, the extra-galactic and Galactic foreground emission must be mitigated, being up to five orders of magnitude brighter than the cosmological 21-cm signal \citep[e.g.][]{jelic_etal:2008,bowman_etal:2009,bernardi_etal:2009,bernardi_etal:2010}. In principle, a separation between the two signals can be made based on the spectral smoothness. The foreground emission, being dominated by synchrotron radiation, is expected to be smooth over tens of MHz, while the 21-cm signal, being a hyperfine spectral line, fluctuates rapidly over MHz-scales \citep{shaver_etal:1999,morales_hewitt:2004,santos_etal:2005,morales_wyithe:2010}. However, instrumental systematics and incorrect calibration can cause spatial and spectral fluctuations in the observed data and mix the signals, making it harder to separate them \citep[e.g.][]{datta_etal:2010,vedantham_etal:2012,barry_etal:2016,ewall-wice_etal:2017,mazumder_etal:2022,gorce_etal:2023}. For this reason, having a precise model of the sky emission, with correct spectral information, is vital to minimise such errors. 

Foreground avoidance methods, such as the ones used by the HERA collaboration, are less dependent on sky modelling, because they mitigate the foreground by using the foreground avoidance technique, which discards the Fourier modes dominated by the foregrounds \citep[e.g.][]{parsons_etal:2012,thyagarajan_etal:2015}. However, they might still need accurate sky models to improve the calibration \citep[e.g.][]{li_etal:2018,kern_etal:2020} or to access the most sensitive small Fourier modes \citep[e.g.][]{kerrigan_etal:2018,ghosh_etal:2020}. On the other hand, sky-modelling subtraction methods, such as the ones used by the LOFAR team, require an accurate sky model \citep{mertens_etal:2020,gan_etal:2022}. The latest upper limits published by the LOFAR-EoR Key Science Project (KSP) showed an excess variance that, on the larger angular scales, reaches 22 times the thermal noise in the residual power of the North Celestial Pole (NCP) field after foreground removal \citep{mertens_etal:2020}. There might be multiple causes for this excess power, such as polarisation leakage \citep{jelic_etal:2015,asad_etal:2015,asad_etal:2016}, model incompleteness \citep{barry_etal:2016, patil_etal:2016, ewall-wice_etal:2017}, low-level radio frequency interference \citep[RFI;][]{wilensky_etal:2019,offringa_etal:2019b}, and uncorrected ionospheric effects \citep{vedantham_koopmans:2016}. However, \citet{gan_etal:2022} concluded that the excess seen in LOFAR data is likely dominated by residuals of bright sources outside the main lobe of the primary beam. The sources that mainly affect the NCP field are Cas\,A and Cyg\,A, which are approximately $30^\circ$ and $50^\circ$ away from the field centre, respectively. At that distance, the chromatic effects of the primary beam are stronger and can introduce rapid spectral fluctuations to the sources, mixing their emission with other foregrounds and the 21-cm signal \citep{pober_etal:2016}. If the beam sidelobes have large spatial and spectral gradients and the source is extended enough, the beam could also imprint different spectral structures to different parts of such a source, making it even harder to calibrate for it \citep{cook_etal:2022}. 

Previous efforts to model bright and extended galaxies have been made, for example, by \citet{procopio_etal:2017} for MWA observations. They demonstrated that when such sources are incorrectly modelled as point sources, their subtraction leaves strong residuals. As a result, an improved wide-field model was made by \citet{lynch_etal:2021} for use in the MWA processing pipeline. Efforts to improve models of bright sources were also made by \citet{line_etal:2020} on the Fornax\,A radio galaxy. In this case, shapelets \citep{refregier:2003} were employed to model the extended emission. Shapelets allow a source to be modelled with fewer components than using a collection of point sources and 2D Gaussian functions, thereby reducing the computational time required for calibration. However, incorporating spatially varying spectral information into shapelets remains challenging and has not yet been tested in real scenarios. \citet{line_etal:2020} showed that, while a theoretical improvement in the 21-cm power spectrum was achieved with the shapelet model, the improvement was negligible in real observations due to other systematics. Similar results were observed by \citet{gehlot_etal:2022} when modelling diffuse Galactic emission in AARTFAAC observations. Shapelets have also been applied to LOFAR observations by \citet{yatawatta_etal:2013}, who made a model of 3C\,61.1, the brightest source in the NCP field. This model was subsequently used by \citet{mertens_etal:2020}. However, the fixed spectrum of the shapelet components proved to be a limitation, and Mertens et al.\ (in prep.) showed that using a multi-scale model of 3C\,61.1 with a spatially varying spectral index led to improved upper limits. \citet{mckinley_etal:2022} showed that combining point sources and Gaussian components can still produce an accurate, high-resolution model of a large source such as Centaurus A. All these efforts suggest that having a spatially and spectrally realistic model of A-team sources is a fundamental requirement to further investigate the cause of the excess power.

In this paper, we will focus on Cyg\,A, which is the brightest A-team source at 150\,MHz. It is a complex source, with both extended and compact  emission. At low-frequencies, its spectral energy distribution deviates from a pure power law. Using several measurements between 12\,MHz and 15\,GHz, \citeauthor{mckean_etal:2016} (\citeyear{mckean_etal:2016}; henceforth cited as MK16) showed that the total flux density $S$ at frequency $\nu$ is well described by a third-order ($n=4$) logarithmic polynomial function of the form
\begin{equation}\label{eq:logpol}
    S(\nu) = S(\nu_0) \prod_{i=1}^{n-1} 10 ^{c_i\log_{10}^i\left(\nu/\nu_0\right)}\, ,
\end{equation}
where $c_1$ corresponds to the spectral index $\alpha$ and $c_2$ to the spectral curvature $\beta$ (see \mksixteent\ for the coefficients values). This total spectral shape does not describe the spatial variation of the Cyg\,A emission, whose spectral indices go from $\alpha\approx -0.65$ in the lobes to $\alpha\approx-1.25$ in the plume region, with a rapid turnover in the two brightest hotspots, the western of which is labelled A and the eastern D (\citealt{leahy_etal:1989}; \mksixteent). In particular, the spectral index of hotspot A, measured at 150\,MHz, is $\alpha_\text{A}=+0.18\pm0.01$, with the turnover around 140\,MHz, while for hotspot D it is $\alpha_\text{D}=+0.36\pm0.02$, with the turnover at 160\,MHz \mksixteenp. 

The main goal of this work is to create an accurate low-frequency spatial and spectral model of Cyg\,A and assess its impact on the LOFAR 21-cm power spectrum. Both \citet{mertens_etal:2020} and \citet{gan_etal:2022} used models of Cyg\,A with a spatially-fixed spectral index $\alpha = -0.8$ at a resolution that is not enough to capture the finest structures of the radio galaxy. By using a better spatial and spectral model of Cyg\,A, we can investigate whether an improved model of a bright off-axis source reduces the observed excess power in the 21-cm signal power spectra. In order to embed physical spectral information into the Cyg\,A model, we take advantage of the forced-spectrum method during a multi-frequency deconvolution \citet{ceccotti_etal:2023}. This method uses an input spectral index map to constrain the spectral index of each component by fitting a logarithmic polynomial function during the deconvolution. Without this regularisation of the spectral behaviour, logarithmic (or ordinary) polynomial fitting can generate unrealistic spectral indices, especially when the frequency bandwidth is limited or systematics in data are high \citep{offringa_etal:2016}. In this work, we used an updated version of the method that supports higher order terms, such as the curvature, as long as maps of the higher order terms are also provided. The spectrally-varying model of Cyg\,A that we get from the forced-spectrum method is the first of its kind in the analysis of the LOFAR power spectrum.\footnote{Main field sources have different spectral indices, but they are only described by single point-like components, so there is no spectral variation within a single source.} Another important goal of this work is to provide a strategy to make a model with physical spectral information by using the forced-spectrum method, thereby making it easier to make accurate source models. The final aim is given by combining the two goals: assessing the importance of improved models of off-axis sources, including spectral information, and the role they play in the excess power that is observed in the 21-cm power spectra.

In Sect.~\ref{sec:cyga-model}, we describe how we obtained the new Cyg\,A model, and compare it with the model used in the current LOFAR-EoR processing pipeline. In Sect.~\ref{sec:ncp-data sets}, we introduce both the simulated and observed NCP data sets that we use to test the impact of the new model onto the 21-cm upper limits. In Sect.~\ref{sec:ncp-processing}, we describe the NCP data processing, from calibration to power spectrum estimation. The results on the two data sets are presented in Sect.~\ref{sec:results}. Finally, in Sect.~\ref{sec:conclusions}, we summarise and discuss the results, drawing the conclusions.  

\section{Cygnus\texorpdfstring{\,}{}A modelling with LOFAR HBA}\label{sec:cyga-model}

The current LOFAR-EoR processing pipeline uses the Cyg\,A model obtained from low-frequency VLA observations\footnote{\url{https://github.com/lofar-astron/prefactor/tree/master/skymodels}} (see Appendix~\ref{app:cyga-old-model} for the catalogue of the model components), which we will refer to as the `old' Cyg\,A model. This model has been used by standard LOFAR processing pipelines, such as \textsc{linc}\footnote{\url{https://linc.readthedocs.io/en/latest/}} \citep[LOFAR Initial Calibration pipeline;][]{degasperin_etal:2019}, for calibration or demixing. Here we identify three major issues of this model when applied to HBA data:
\begin{enumerate}[i.]
    \item \emph{Spectral extrapolation}: the flux density of each component has been derived from a clean-based model of VLA observations at 73.8\,MHz, and it is then extrapolated to higher frequencies using a given spectral index $\alpha$;
    \item \emph{Low resolution}: the model consists of four point and eight Gaussian components, five of which have a zero-length minor axis, that are unable to model the complex structures of the source visible at higher frequencies (see Sect.~\ref{sec:cyg-model-comparison});
    \item \emph{Spatially constant spectral index}: each component has an artificial, constant spectral index $\alpha=-0.8$, even though the source has a strongly varying spectral index. 
\end{enumerate}

The total flux density given by the old Cyg\,A model at 73.8\,MHz is 15.9\,kJy, while the results of \mksixteent\ indicate it is $16.3 \pm 2.2\,\text{kJy}$ when extrapolated to that  frequency using Eq.~\eqref{eq:logpol}. Whereas this offset is acceptable within the $1\sigma$ error, issues (i) and (iii) lead to a much larger difference at higher frequencies. Extrapolating the flux density of the old model to 150\,MHz by using a first-order ($n=2$) logarithmic polynomial with $\alpha=-0.8$ results in a total flux of 9.0\,kJy, which is 16\% lower than the expected 10.7\,kJy \mksixteenp. The inconsistent flux density scale is another reason for improving the model of Cyg\,A. We will describe the new model in the next section. 

\subsection{The new Cygnus\texorpdfstring{\,}{}A model}\label{sec:new-model}

LOFAR has the capabilities to operate between 110 and 250\,MHz, by selecting two sub-ranges of the high band antenna (HBA) system, HBA-low in the 110--190\,MHz range and HBA-high in the 210--250\,MHz range. For each observation, a total of 488 sub-bands (SB) can be recorded, each with 195.3\,kHz of bandwidth, which are further divided into 64 frequency channels. LOFAR currently consists of 24 core stations (CS) near the centre of the array, distributed within a diameter of ${\approx}4$\,km and 14 remote stations (RS) across the Netherlands, with a maximum baseline of ${\approx}120$\,km, forming what is called the Dutch array. Another 14 (international) stations are spread across Europe, providing baselines up to ${\approx}2000$\,km. Each CS includes two HBA sub-stations (HBA0 and HBA1) that can either function independently (`HBA Dual' mode), improving the $uv$-coverage, or be individually selected (`HBA Zero/One' mode).

Given the large frequency coverage of the HBA system and the high spatial resolution provided by the Dutch array (approximately 3\,arcsec), our goals with the improvement of the model for Cyg\,A are:
\begin{enumerate}[i.]
    \item \textit{Obtain an accurate model at 120--160\,MHz}: this is the frequency range targeted by the LOFAR-EoR KSP and, therefore, we want to use HBA observations;
    \item \textit{Increase resolution to 5\,arcsec}: hotspots are compact sources and we need a sufficient high resolution to not mix their emission with the surrounding lobes: 5\,arcsec is the highest resolution provided by LOFAR HBA to get a model that is accurate for all Dutch baselines;
    \item \textit{Capture the spatially varying spectral index}: Cyg\,A is a morphologically complex radio galaxy, where different regions show very different spectral behaviours that have to be taken into account.
\end{enumerate}

\begin{table}
\caption{Observational details of the Cyg\,A data set after flagging, averaging, and calibration.}
\label{tab:obs-details-cyga}
\centering
\begin{tabular}{lc}
\toprule
Parameter & Value \\
\midrule
Telescope & LOFAR HBA \\
Project code & LC1\_013 \\
Antenna configuration & HBA One \\
Number of stations & $36$ ($\text{CS} + \text{RS})^a$  \\
Phase centre (J2000): & {} \\
{ }{ } Right Ascension & $19^\text{h}59^\text{m}28\rlap{.}^\text{s}00$\\
{ }{ } Declination & $+40^\circ44'02\rlap{.}''00$ \\
Obs.\ start time (UTC): & {} \\
{ }{ } HBA-low & 2014 May 13; 00:11 \\
{ }{ } HBA-high & 2014 May 14; 00:00 \\
Frequency range: & {} \\
{ }{ } HBA-low & 110.7--181.4\,MHz \\
{ }{ } HBA-high & 210.0--248.6\,MHz \\
Duration of observation & 12\,h  \\
Time resolution & 6.0\,s \\
Frequency resolution & 195.3\,kHz\\
\bottomrule
\end{tabular}
\tablefoot{
\tablefoottext{a}{CS006 and RS208 did not participate in this observation.}
}
\end{table}

To satisfy goal (i), we use a combination of 12\,h LOFAR HBA-low and HBA-high observations taken in 2014. The data, previously processed and calibrated (see Appendix~\ref{app:cyga-processing}), consist of 412 SBs spanning 111--181 and 210--249\,MHz, with a frequency resolution of 195.3\,kHz (i.e.\ one channel per SB) and an integration time of 6\,s. Other observational details are reported in Table~\ref{tab:obs-details-cyga}. All modelling and data processing described in this paper has been performed on the `Dawn' high-performance computing cluster \citep{pandey_etal:2020} at the University of Groningen.

In the next sections, we describe how we achieved goals (ii) and (iii) to get a new, more complete model of Cyg\,A. We used the \textsc{wsclean} imager software \citep{offringa_etal:2014} to make images of the data. We start by describing the deconvolution methods that we employed. 

\subsection{MS-MF deconvolution and forced-spectrum fitting}

Modelling the spectral structures of a radio source requires fitting some smooth function to the data from the frequency channels. This can be done during a multi-frequency (MF) clean approach, which models the spectral behaviour during deconvolution \citep{sault_wieringa:1994}. In the joined-channel deconvolution described by \citet{offringa_smirnov:2017} and implemented in \textsc{wsclean}, a wide frequency band is divided into narrower output channels which are jointly imaged and deconvolved, and combined into a high dynamic range, frequency-integrated continuum image. The integrated image is used by the cleaning process to locate the peak of each model component. The frequency-varying brightness is measured across the images at each output frequency channel. This MF approach allows the components to capture the spectral variations of the source. Fitting the measured spectrum of each found component to an ordinary or logarithmic polynomial function ensures spectral smoothness of the model, which can be expected for sources dominated by synchrotron emission \citep[e.g.][]{reich_reich:1988,tegmark_etal:2000,jackson:2005,jelic_etal:2008}.

However, a clean-component spectral model might contain incorrect spectral information because, for instance, of the limited frequency bandwidth or when systematics in the data are large \citep[e.g.][]{rau_etal:2016,offringa_etal:2016}. As a result, such spectral indices can not be used for extrapolation. The forced-spectrum method \citep{ceccotti_etal:2023} solves this problem by using a spectral index map to assign to each clean-component a specific spectral index value based on the position of the component, fitting a logarithmic polynomial function of the form of Eq.~\eqref{eq:logpol}. A modified weighted linear least-squares is then used to only determine the magnitude $S(\nu_0)$ of each clean-component. \citet{ceccotti_etal:2023} used the forced-spectrum method to produce a model with only spectral indices. In this work, we additionally constrain higher order terms, such as the spectral curvature. This is necessary to model the spectral turnover of the Cyg\,A hotspots. When using higher order terms, it is important that the input spectral maps have been calculated at the same reference frequency at which \textsc{wsclean} performs the MF spectral fitting. In Sect.~\ref{sec:new-model-imaging} we will show a method to make spectral index and curvature maps from in-band data.

The forced-spectrum method, and in general the MF deconvolution, works well in combination with a multi-scale (MS) method \citep{cornwell:2008,rich_etal:2008,rau_cornwell:2011}. The MS deconvolution can incorporate diffuse emission by making a cleaned model as a summation of delta (i.e.\ point-like components) and basis functions (e.g.\ Gaussian or tapered quadratic). In this work, we used the MS deconvolution implemented in \textsc{wsclean} \citep{offringa_smirnov:2017} to model the extended emission with (circular) Gaussian functions that have an analytically defined Fourier transform, unlike tapered quadratic functions used by the algorithm described by \citet{cornwell:2008}. The combination of the MS and forced-spectrum methods result in a compactly described model, which is essential for calibration.

\subsection{Imaging and spectral modelling of Cygnus\texorpdfstring{\,}{}A}\label{sec:new-model-imaging}

To build the high-resolution spectral model of Cyg\,A, our imaging process was divided into four key steps: initial high-resolution imaging, imaging for accurate spectral modelling, spectral index extraction, and forced-spectrum fitting. This approach was designed to first capture the spatial details of the source, and then to accurately translate these observations into a comprehensive spectral model. The first step, the initial imaging, is optional and was done to check the data quality and the spectral features of hotspots A and D at the highest resolution allowed by the data set.  

\subsubsection{High-resolution imaging and spectra of the hotspots}\label{sec:high-res-imaging}

Our Cyg\,A data set spans a wide bandwidth, from approximately 110 to 250\,MHz, allowing us to create images with a synthesised beam (i.e.\ the point-spread function) of about 2\,arcsec in size at the highest frequency. Although we can not take advantage of this for reliable spectral index extraction, where maintaining a consistent beam size across the bandwidth is essential to ensure each frequency samples the same spatial scale, we can utilise it to generate high-resolution images for data quality assessment. However, if we clean deep enough so that the residual is mostly free of emission from the source, we can set the restoring beam to a fixed value and still get good spectral information from the restored images. This is the approach used in this section to extract the spectral behaviour of the hotspots. 

For this initial step, we used the MF joined-channel deconvolution, splitting the full bandwidth into 100 independent (i.e.\ without spectral fitting) output channels. We did not use MF weighting, ensuring a better image quality for each output frequency by gridding the $uv$-plane of each channel on a separate grid. The output channels were weighted with a Briggs weighting scheme with a robust parameter of $-2$, resulting in an integrated synthesised beam with a full width at half-maximum (FWHM) of $3\times2$\,arcsec. We used the MS deconvolution with point and Gaussian components of size 5, 10, 15, 20, 40, and 80\,arcsec. We performed the cleaning to an initial threshold of $5\sigma$ using auto-masking and a final threshold of $1\sigma$, where $\sigma=8.3\,\text{mJy/beam}$ for the frequency integrated image, with a dynamic range of $1:19\,540$. We found that using a $1\sigma$ cleaning threshold, the residuals of the 100 output channels are noise-like at the hotpots location, but not completely on the rest of the source. This produced 10 times stronger residuals  at the edges of the source at low frequencies and in the lobes at high frequencies. These are probably deconvolution artefacts caused by to the morphological complexity of the source. Because the condition of having the residual free of source emission is met for the brightest structures of Cyg\,A, we can still set a constant restoring beam to extract the spectra of hotspots A and D. We used a restoring circular beam with a FWHM of 3\,arcsec, slightly larger than the frequency-integrated synthesised beam. Other imaging parameters are listed in the `High resolution' column of Table~\ref{tab:wsclean-par}. 

\begin{table}
\caption{The \textsc{wsclean} parameters used for the initial high-resolution imaging and the forced-spectrum method, which produces the final model of Cyg\,A.}
\label{tab:wsclean-par}
\centering
\begin{tabular}{lcc}
\toprule
Parameter & High resolution & Forced spectrum \\
\midrule
Image size & $1024\times1024$& $256\times256$ \\
Pixel scale & 0.25\,arcsec& 1\,arcsec \\
Weighting & Briggs $-2$ & Briggs $-2$ \\
Gridder & W-gridder & W-gridder \\
Minor gain & 0.01 & 0.01 \\
Major gain & 0.8 & 0.8 \\
Spatial mask & -- & 1\,Jy/beam [--] \\ 
Auto-mask$^a$ & $5\sigma$ & $3\sigma$ [$6\sigma$] \\
Auto-threshold$^b$ & $1\sigma$ & $1\sigma$ \\
MS scales (arcsec) & 0--80 & 0--99 (default) \\
MS gain$^c$ & 0.2 & 0.05 \\
Output channels & 100 (joined) & 41 (joined) \\
Spectral fitting & -- & log.\ pol.\ $n=3$ [--] \\
$uv$-range (k$\lambda$) & -- & $0.16{-}45.5$ \\
Beam size & 3\,arcsec &  5\,arcsec \\
\bottomrule
\end{tabular}
\tablefoot{
Most of the parameters from the forced-spectrum method were also applied when generating the images used to extract the spectral index and curvature maps. The specific parameters for this imaging process are provided in square brackets.\\
\tablefoottext{a}{Initial threshold at which a mask is constructed by finding clean-components, based on the noise standard deviation.}
\tablefoottext{b}{Stopping threshold inside the mask, based on the noise standard deviation.}
\tablefoottext{c}{Size of step made in the MS sub-minor loop.}
}
\end{table} 

\begin{figure*}
    \centering
    \includegraphics[width=1\textwidth]{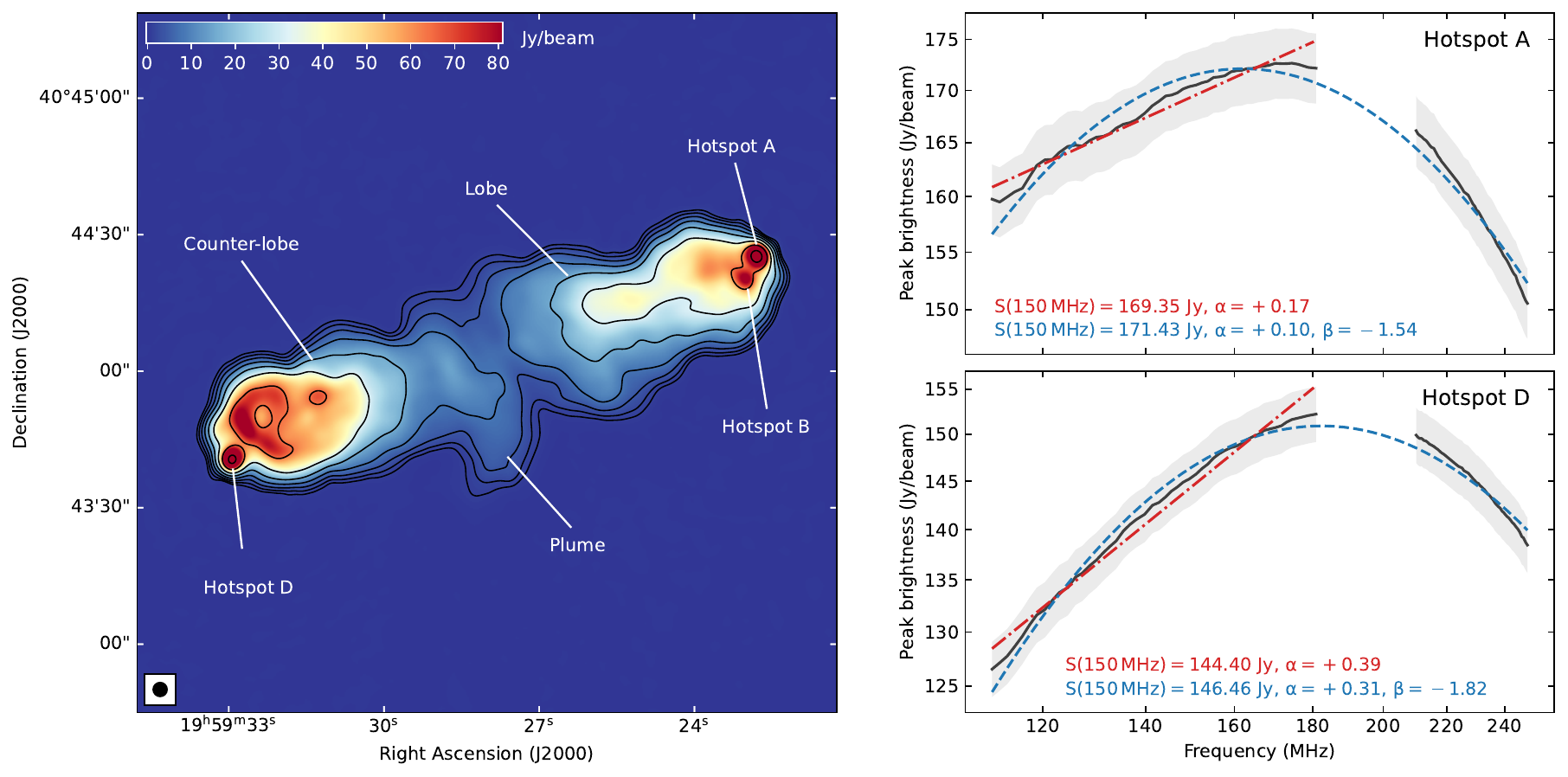}
    \caption{High-resolution, frequency-integrated image of Cyg\,A between 111 and 249\,MHz (left), and peak brightness of the hotspots A (top right) and D (bottom right), extracted from the high-resolution images of the 100 output channels (black solid line, with $1\sigma$ uncertainties shown as a grey shaded area). The fits using a second-order (blue dashed line) and a first-order logarithmic polynomial function (red dot-dashed line) are also shown, with the best fit parameters reported in the same colours. The latter has been fitted only between 111 and 181\,MHz for a consistent comparison with \mksixteent. The image on the left has a noise level of $\sigma=10\,\text{mJy/beam}$, with contours starting at $1\,\text{Jy/beam}$ and increasing by a factor of 2.}
    \label{fig:CygA_imageHR_hotspots_peak}
\end{figure*}

\begin{figure*}
    \centering
    \includegraphics[width=1\textwidth]{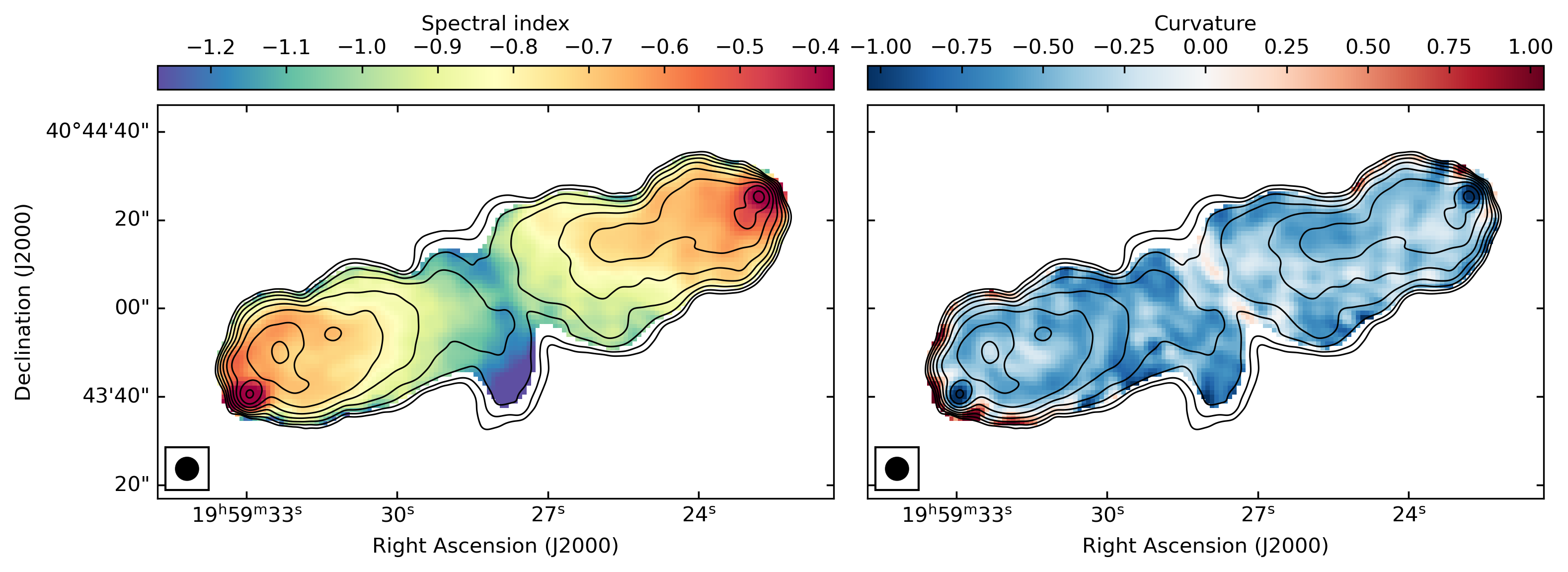}
    \caption{Forced-spectrum input spectral index (left) and curvature maps (right) at 190.18\,MHz, extracted from the 5\,arcsec resolution images. Contours in both figures are the same of the left panel of Fig.~\ref{fig:CygA_imageHR_hotspots_peak}. The colour scale of the spectral index map spans the range $-1.27\le \alpha \le -0.38$, while the curvature is plotted in the range $-1.04\le \beta \le 1.04$.}
    \label{fig:CygA_new_spix_curv}
\end{figure*}

The resulting 111--249-MHz frequency-integrated image is shown in the left panel of Fig.~\ref{fig:CygA_imageHR_hotspots_peak}. At 3\,arcsec resolution, the two brighter hotspots, A and D, are well resolved and we can extract their spectra from the 100 output images. We fitted their peak brightness using a non-linear least-squares method with a second-order ($n=3$) logarithmic polynomial of the form
\begin{equation}\label{eq:logpol_3terms}
    S(\nu) = S(\nu_0)\left(\frac{\nu}{\nu_0}\right)^{\alpha} 10^{\,\beta \log_{10}^2 (\nu/\nu_0)}\, , 
\end{equation}
which is equivalent to Eq.~\eqref{eq:logpol} when using three terms, where $\alpha$ and $\beta$ are, respectively, the spectral index and the curvature terms evaluated at the reference frequency $\nu_0$. We used 150\,MHz as reference frequency to compare our spectral values with \mksixteent. The panels on the right-hand side of Fig.~\ref{fig:CygA_imageHR_hotspots_peak} show these fits with a blue dashed line. Here, the black solid line represents the peak brightness as extracted from each output image, to which we have associated $1\sigma$ errors (grey shaded area). These are calculated by adding in quadrature the standard deviation of the noise of each image and a standard calibration error of 2\% of the flux density \mksixteenp, where the latter dominates the uncertainties. We find that the best fit values for the hotspot A are $\alpha_\text{A} = +0.1025 \pm 0.0002$ and $\beta_\text{A}=-1.541 \pm 0.009$, and for the hotspots D are $\alpha_\text{D} = +0.3081 \pm 0.0002$ and $\beta_\text{D}=-1.819 \pm 0.009$, confirming the presence of a turnover in both of them. Our values differ from those found by \mksixteent\ for a couple of reasons: they only had HBA-low data in the range 109--183\,MHz, and they fitted a first-order logarithmic polynomial. Restricting our analysis in the same frequency range and fitting a simple power law without the curvature term (red dot-dashed line in Fig.~\ref{fig:CygA_imageHR_hotspots_peak}), we find $\alpha_\text{A}=+0.1720 \pm 0.0004$ and $\alpha_\text{D}=+0.3917 \pm 0.0004$, which are consistent with the values reported by \mksixteent, namely, $\alpha_\text{A}=+0.18\pm0.01$ and $\alpha_\text{D}=+0.36\pm0.02$. The $1\sigma$ uncertainties on $\alpha$ and $\beta$ assume uncorrelated errors, whereas systematics such as calibration and deconvolution errors are typically correlated, and are therefore not captured by these uncertainties.

\subsubsection{Spectral index extraction}\label{sec:spix-extraction}

The second step towards constructing an accurate spectral model of Cyg\,A was to generate images that could be used for a pixel-by-pixel spectral index extraction. As mentioned in the previous section, using the 100 high-resolution images was not ideal due to the ${\approx}10\sigma$ residual emission, which could affect the spectral indices of fainter emission. While the clean model would be convolved with a consistent restoring beam across the bandwidth, the residual images retain the synthesised beam of their respective output channels. This might result in a mix of components at different spatial scales from low to high frequencies, preventing the extraction of a reliable source spectrum. To address this, we performed a separate imaging step with appropriate settings for an accurate full-source spectral index extraction.

Because we could not be certain that the residual images were noise-like across the entire bandwidth, we ensured a consistent synthesised beam (i.e.\ the same $uv$-coverage) by applying a strict $uv$-cut. We selected only the (projected) baselines in the range $160{-}45\,500\,\lambda$, where the lower value corresponds to the shortest baseline at 249\,MHz and the upper value to the longest baseline at 111\,MHz. By removing baselines shorter than $160\,\lambda$, we lost sensitivity to structures larger than ${\approx}20\,\text{arcmin}$, which was not a concern since our focus was only on Cyg\,A, which has a maximum extent of 2.3\,arcmin. Conversely, the upper $uv$-limit reduced the resolution from 3 to 4.5\,arcsec, leading us to fix the restoring beam to a circular Gaussian with a FWHM of 5\,arcsec. Consequently, we increased the pixel scale from 0.25\,arcsec (used for high-resolution imaging) to 1\,arcsec, producing images of $256\times256$\,pixels. We also reduced the number of output channels to 41 to get a better noise per output image. This number was still high enough to minimise any bias that would occur during the later fitting process. We measured a noise level of $\sigma=12.8\,\text{mJy/beam}$ for the frequency-integrated image (dynamic range of $1:20\,550$). With these settings, we found that the MS deconvolution gave better results with the default spatial scales of 12, 25, 49, and 99\,arcsec. The other \textsc{wsclean} parameters were similar to those used in the high-resolution imaging. As we will apply similar settings in the next forced-spectrum fitting step, all parameters are listed in the `Forced spectrum' column of Table~\ref{tab:wsclean-par}, with specific parameters for this step provided in square brackets.

The resulting 41 output images were used to extract spectral index and curvature maps. The brightness of each pixel was fitted using Eq.~\eqref{eq:logpol_3terms} at the reference frequency of 190.18\,MHz, which is the same frequency at which \textsc{wsclean} extracted the model clean-components for our specific settings.\footnote{The reference frequency needs to be checked beforehand each time, as different weighting and $uv$-coverage can shift it with the same data set.} In this operation, we considered only pixels brighter than $10\,\text{Jy/beam}$ in the frequency-integrated image (this roughly corresponds to 3.6\,Jy/beam in the frequency-integrated, high-resolution image of Fig.~\ref{fig:CygA_imageHR_hotspots_peak}). The extracted spectral index and curvature maps are shown in Fig.~\ref{fig:CygA_new_spix_curv}. Although the image resolution is lower than in the preliminary imaging step, the large LOFAR HBA bandwidth and high image dynamic range still allow us to identify the main spectral features across the entirety of Cyg\,A. Hotspots A and D exhibit the flattest spectral indices and the lowest curvature, as expected due to their turnover. Additionally, in line with synchrotron spectral ageing models (e.g.\ \citeauthor{carilli_etal:1991} \citeyear{carilli_etal:1991}; \mksixteent), the spectral index map shows a steepening from the hotspots, where relativistic particles are continuously accelerated by the jets, to the outer regions of the lobe and counter-lobe, where the primary energy loss mechanism is synchrotron radiation. More energetic plasma, which emits at higher frequencies, loses energy faster than less energetic plasma, creating a break frequency that shifts to lower frequencies as the plasma ages (unless re-acceleration occurs). Because the plume of Cyg\,A has the steepest spectral index ($\alpha < -1$), it consists of the oldest plasma in the source, indicating that this plasma underwent its last acceleration much earlier than in any other region.

\subsubsection{Forced-spectrum fitting and spectral rescaling}\label{sec:forced-spectrum}

In both Sect.~\ref{sec:high-res-imaging} and \ref{sec:spix-extraction}, the imaging was performed without any constraints on the spectral behaviour. In the next processing step, the extracted spectral index and curvature maps were used as inputs to the forced-spectrum method, acting as an imaging regularisation. We mostly applied the same \textsc{wsclean} settings as in the previous step, as listed in the `Forced spectrum' column of Table~\ref{tab:wsclean-par}. The same $uv$-range was used to ensure that the sampled spatial scales were consistent with the spectral index map. While we could have used higher resolution to obtain a more detailed spatial model of Cyg\,A, the 5\,arcsec resolution of the spectral index map meant that this would not have provided any additional benefit on the spectral modelling side. Furthermore, we preferred to maintain the same $uv$-cut and resolution to limit the number of clean-components in the final model, which consisted of 2082 components (1320 point sources and 762 Gaussians).

Due to the spectral constraints of the forced-spectrum method, the residual images may exhibit higher residuals, resulting from the clean-component overlap described in \citet{ceccotti_etal:2023}. This occurs because the spectral index is assigned based on the clean-component position, but for Gaussians, the wings overlap with other components in regions where the spectral index differs, causing a bias in the final model image and, consequently, in the residual image. For spectrally complex sources such as Cyg\,A, this bias can be significant, particularly in the extended regions where many Gaussians are used during cleaning. To mitigate this, we constrained the cleaning process using a spatial mask that included only pixels brighter than 1\,Jy/beam, based on the previous frequency-integrated image (corresponding to approximately 0.4\,Jy/beam in the high-resolution image of Fig.~\ref{fig:CygA_imageHR_hotspots_peak}). This approach allowed us to push the initial cleaning threshold deeper, reducing residual emission and preventing artefacts from being cleaned beyond the source extension. As a result, clean components may still be selected between the spatial mask boundary and the spectral index map extension, which was defined down to 10\,Jy/beam. To minimise the impact of the overlap bias in the spectrum of the final model within the 10\,Jy/beam contour, we assigned values of $\alpha=0$ and $\beta=0$ to the remaining areas of the spectral maps.

\begin{figure}
    \centering
    \includegraphics[width=1.\columnwidth]{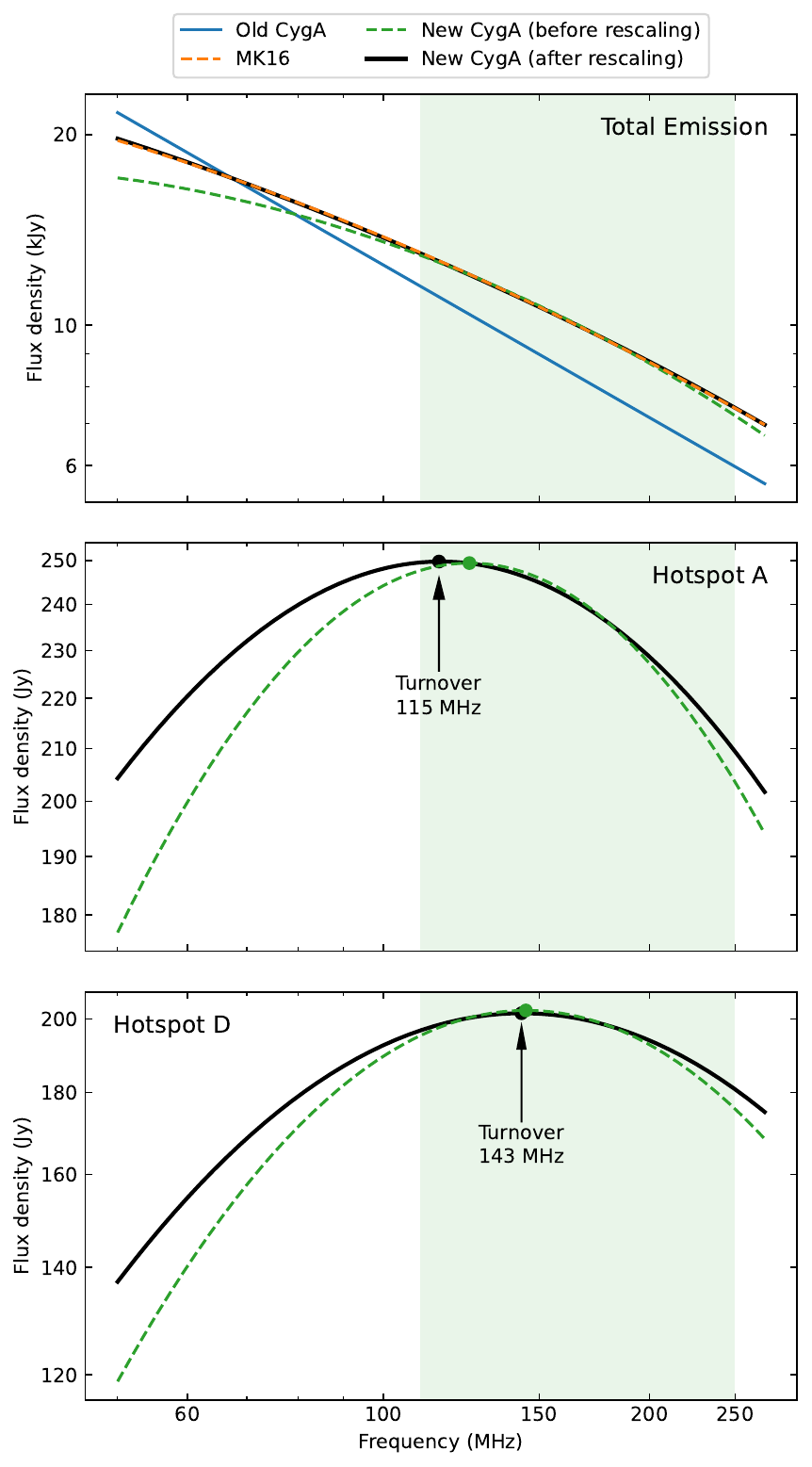}
    \caption{Flux density between 50 and 270\,MHz measured from the Cyg\,A models for the total emission (top), hotspot A (middle), and hotspot D (bottom). The top panel shows the old model (solid blue line), the new model before (dashed green line) and after spectral rescaling (solid black line), and the third-order logarithmic polynomial used by \mksixteent\ (dashed orange line). In the middle and bottom panels, only the flux densities from the new Cyg\,A models before and after rescaling are shown. These values were extracted from a circular region with a 5\,arcsec diameter, centred on the peak of each hotspot. The filled points indicate the spectral peaks, and the turnover frequency after rescaling is also reported. The green shaded area in all panels marks the LOFAR HBA range used to extract the new Cyg\,A model.}
    \label{fig:CygA_model_flux_comp_hs}
\end{figure}

The output of our forced-spectrum imaging step is a new, high-resolution model of Cyg\,A, whose total flux density matches, between 110 and 250\,MHz, the spectrum found by \mksixteent\, as shown in the top panel of Fig.~\ref{fig:CygA_model_flux_comp_hs}. However, when extrapolated outside the LOFAR HBA frequency range, the spectral energy distribution diverges from \mksixteent, likely because a sum of components with varying spectra would not necessarily follow the flux scale to which the data were calibrated. Moreover, the overlap bias might play a role. While extrapolation always requires caution, this new Cyg\,A model could be useful for applications at lower frequencies, such as calibrating or demixing LOFAR low band antenna (LBA) data, between 10 and 80\,MHz. To address this, we evaluated and corrected the total flux density of our model at 10 frequencies between 10 and 300\,MHz to match \mksixteent. Each clean component was then adjusted by the found correction factors and fitted to a logarithmic polynomial to restore spectral index and curvature values at the model reference frequency. The rescaled model agrees with \mksixteent\ as shown by the black line in Fig.~\ref{fig:CygA_model_flux_comp_hs}. The middle and bottom panels show the flux density of hotspots A and D, integrated within a circle of 5\,arcsec diameter. Compared to Fig.~\ref{fig:CygA_model_flux_comp_hs}, the hotspot spectra appear flatter, with the turnover shifted to lower frequencies. This is due to the lower resolution used for both the spectral maps and the final model, allowing more lobe emission to contaminate the hotspots. Such flattening is accentuated by the rescaling, although most effects impact frequencies outside the LOFAR HBA range.

\subsubsection{Old vs new Cygnus\texorpdfstring{\,}{}A model}\label{sec:cyg-model-comparison}

\begin{figure}
    \centering
    \includegraphics[width=1\columnwidth]{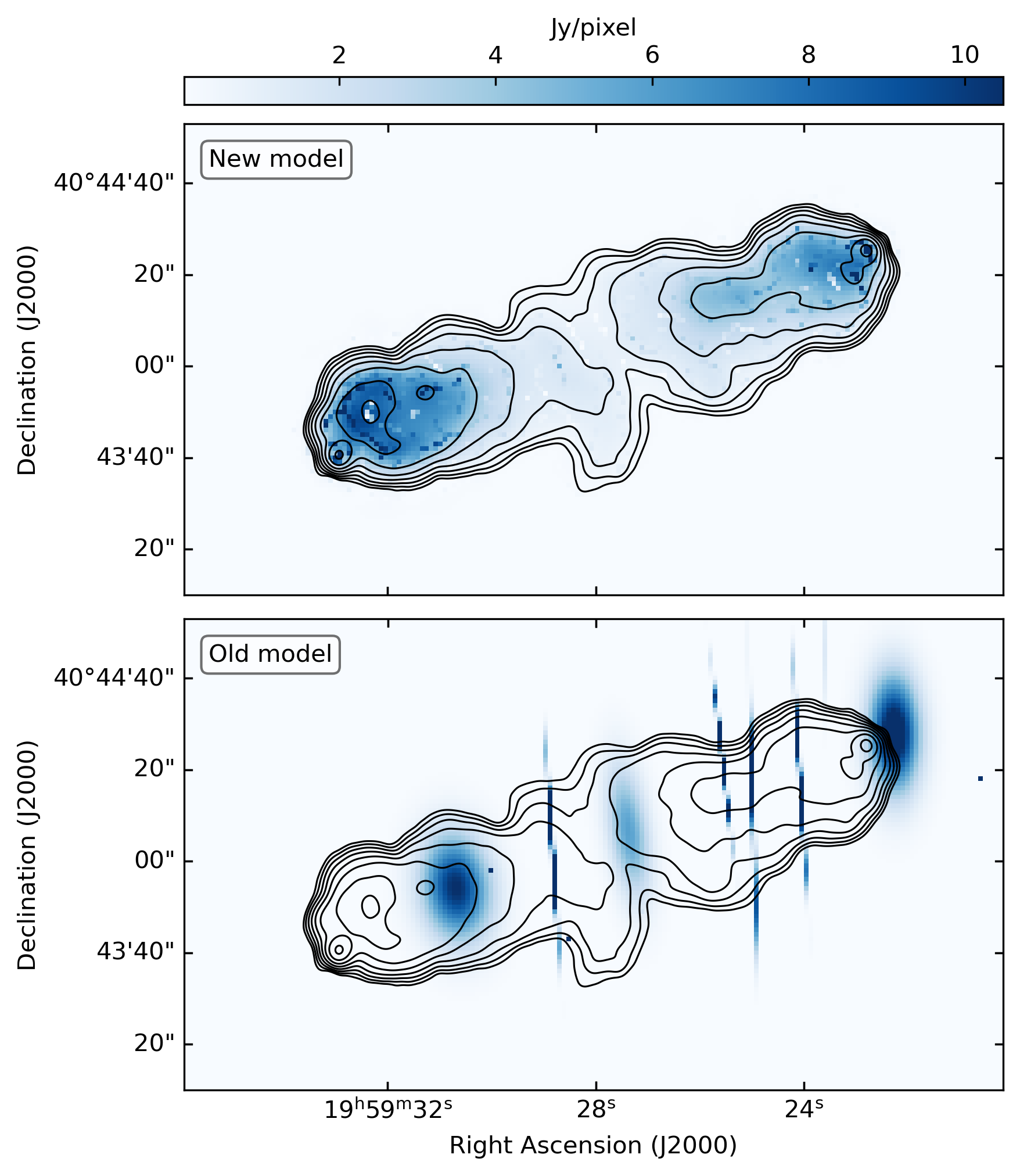}
    \caption{New (top) and old model (bottom) of Cyg\,A, rendered at 150\,MHz. Contours are the same of Fig.~\ref{fig:CygA_imageHR_hotspots_peak}. The zero-length axes of the old model Gaussians have been replaced with 1\,arcsec axes for imaging purposes, resulting in the appearance of vertical spikes.}
    \label{fig:CygA_model_comp}
\end{figure}

The new, rescaled model of Cyg\,A and the old model are shown, respectively, in the top and bottom panels of Fig.~\ref{fig:CygA_model_comp}, both rendered at 150\,MHz. It is immediately clear that the new model is at a higher resolution than the old one. Hotspots A, B and D are modelled with a dominant point component, while the diffuse lobe and counter-lobe emission consists of a collection of Gaussian components of different sizes. Because \textsc{wsclean} uses only circular Gaussians during the MS deconvolution, point components are required to shape these Gaussians into real source structures, such as the bow-shaped structure in the counter-lobe. This explains the presence of components with negative fluxes, that have no physical meaning but they are still needed to get models resembling real data. The final result is a model where also the fine structures of Cyg\,A are visible and easily distinguishable, fulfilling goal (ii), mentioned in Sect.~\ref{sec:new-model}. On the other hand, we can hardly distinguish the different structures in the old model. Assuming that the two outermost Gaussian components model the hotspots A and D emission, a shift in the coordinates toward north-west is present in the old model, the cause of which is unknown. This issue would be easily solved by shifting back each components to the correct location in the sky, but it is hard to find a reference point, such as hotspot A, in this model. When the old model is used on its own for calibration or demixing, such an offset, which is not too large, is absorbed into the gain phases. 

The new model also outperforms the old one in terms of spectral modelling. The forced-spectrum method transferred the pixel-by-pixel spectral information from the spectral index and curvature maps into the new model, whose components have different spectra depending on their location, as we saw in the middle and bottom panels of Fig.~\ref{fig:CygA_model_flux_comp_hs}. Thus, the resulting model of Cyg\,A also satisfies goal (iii). Plotting the old model in the top panel, extrapolated from 74.8\,MHz to the HBA frequency range, shows the large difference in flux scale between the expected values, as already discussed at the beginning of Sect.~\ref{sec:cyga-model}. 


\section{NCP data sets}\label{sec:ncp-data sets}

\begin{table}
\caption{Observational details of the NCP data set used for testing the Cyg\,A model.}
\label{tab:obs-details-ncp}
\centering
\begin{tabular}{lc}
\toprule
Parameter & Value \\
\midrule
Telescope & LOFAR HBA \\
Project code & LC2\_019 \\
Observation ID & L246309 \\
Antenna configuration & HBA Dual Inner \\
Number of stations & $62$ ($\text{CS} + \text{RS})$  \\
Phase centre (J2000): & NCP ($0^\text{h}$, $+90^\circ$)\\
Obs.\ start time (UTC): & 2014 Oct 16; 17:11:36 \\
Frequency range: & 134.1--147.1\,MHz \\
Duration of observation & 12.6\,h [6\,h] \\
Time resolution & 2.0\,s \\
Frequency resolution & 61.0\,kHz\\
\bottomrule
\end{tabular}
\tablefoot{The duration value within square brackets refers to the simulation data set.}
\end{table}

To test the performances of the new model of Cyg\,A against the old one, we used it within the LOFAR-EoR processing pipeline, comparing the results both from a simulated data set and from a real observation of the North Celestial Pole (NCP) field. This allows us to asses the impact of this better model on the 21-cm power spectrum measurements, which is one of the main motivations of this work. We selected an average-quality night from \citet{mertens_etal:2020} where Cyg\,A is above an elevation of ${\approx}10^\circ$ for most of the observing duration. The observational details of the selected data set are reported in Table~\ref{tab:obs-details-ncp}. 

\begin{figure*}
    \centering
    \includegraphics[width=1\textwidth]{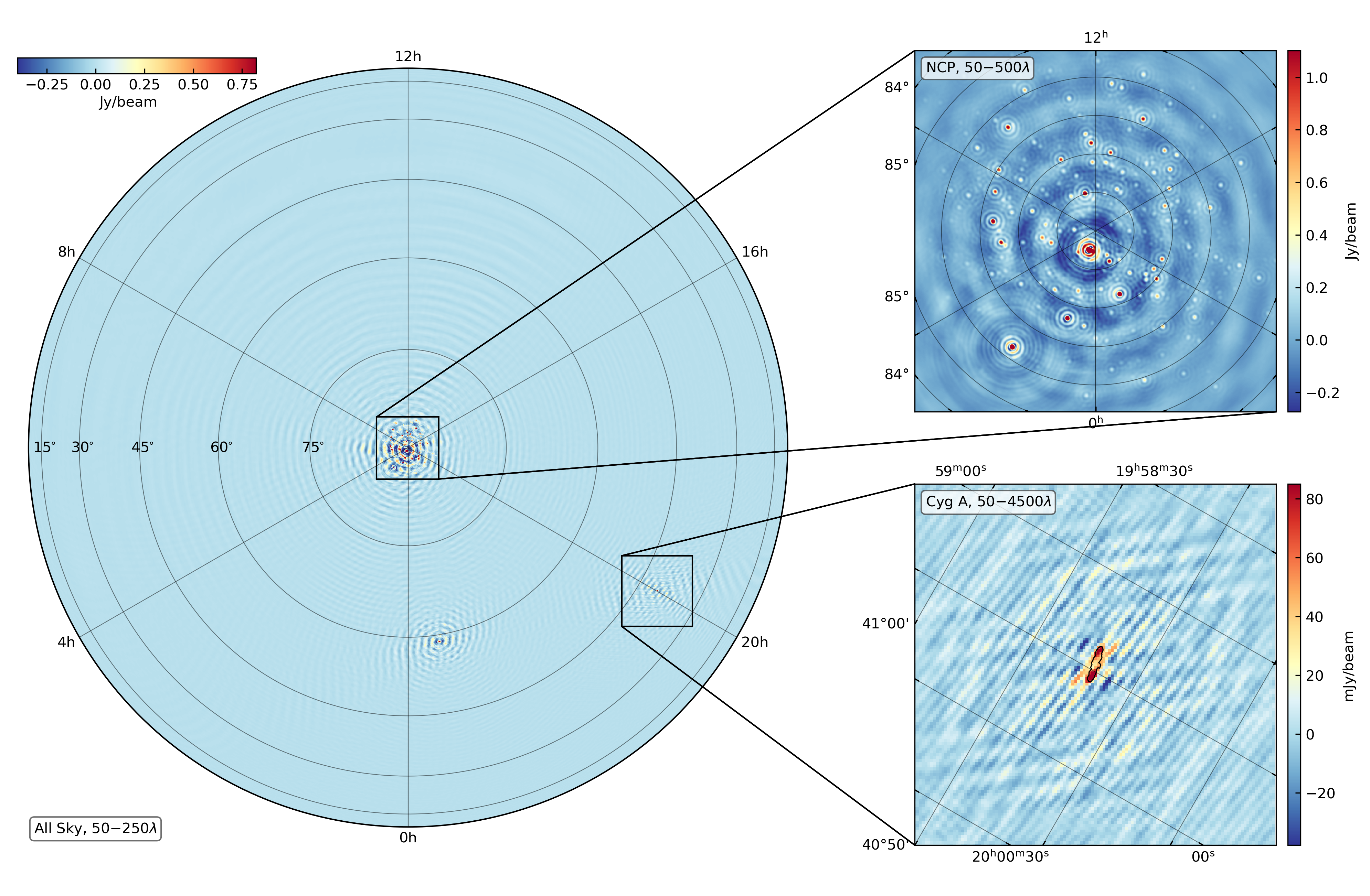}
    \caption{All sky dirty image of the simulated NCP data set (left), with zoom-in to the NCP main field (top right) and the Cyg\,A direction (bottom right). The all sky image has been obtained using a baseline range of $50{-}250\lambda$, the NCP zoom-in with $50{-}500\lambda$, while the Cyg\,A zoom-in with $50{-}4500\lambda$. A natural weighting scheme has been used for the all sky and the NCP zoom-in images, whereas a uniform weighting has been chosen for the Cyg\,A zoom-in to make the source structures more visible. The contour in the bottom right panel is the 1\,Jy/beam level from Fig.~\ref{fig:CygA_imageHR_hotspots_peak} to show the position and extension of Cyg\,A.}
    \label{fig:NCP_sim}
\end{figure*}

To produce the simulated data set, we made a copy of the same measurement set and forward predicted (i.e.\ Fourier transformed from image space to visibility space) a sky model into it. When predicting visibilities, the calculation of the primary beam is computationally expensive. The computing time increases with the number of model components, because the primary beam has to be calculated at each component coordinate. Because the NCP sky model has more than 28\,700 components \citep{yatawatta_etal:2013,mertens_etal:2020}, we used a simpler NCP model with only 684 point components with a flat spectrum, presented in \citet{brackenhoff_etal:2024}. This model extends up to $6^\circ$ from the NCP and consists of 14 source clusters (i.e.\ directions), and its sources dominate the total power spectrum of the sky, making it sufficient for the purpose of this work. More details are reported in Appendix~\ref{app:simple-NCP_model}. We added a low resolution model of Cas\,A, consisting of 24 Gaussian components with $\alpha = -0.8$ constant across the source and $\nu_0=74\,\text{MHz}$, similar to the old Cyg\,A model. Finally, we add our new Cyg\,A model. Assuming that this new model is the true sky, we will be able to measure the theoretical excess power that might be caused by not having a Cyg\,A model with correct spectral indices. 

The combined sky model was predicted on to the visibilities of the previously selected night with \textsc{sagecal}\footnote{\url{https://sagecal.sourceforge.net/}} \citep{yatawatta:2016}, which allows using GPUs to speed up the process (see Sect.~\ref{sec:ncp-processing}). The primary beam was applied during the prediction, and we finally added a per-visibility noise estimated from the radiometer equation \citep[e.g.][]{thompson_etal:2017}:
\begin{equation}\label{eq:radiometer_equation}
\sigma_\text{vis} = \frac{\text{SEFD}}{\sqrt{2 \Delta t \Delta \nu}}\, ,
\end{equation}
where SEFD is the system equivalent flux density, $\Delta t = 2\,\text{s}$ is the visibility integration time and $\Delta\nu = 61\,\text{kHz}$ is the frequency channel width. The SEFD of CS and RS is expected to be different because RS have double collecting area \citep{vanhaarlem_etal:2013}. However, our observations were carried with the `Inner' setup, which sets the RS collecting area to the CS one by switching off the outer tiles of the station. For this reason, we assigned to every visibility the same SEFD of $4253\,\text{Jy}$ as estimated by \citet{mertens_etal:2020}. To reduce the computing time, we kept only the first 6\,h of the data set, where the elevation of Cyg\,A peaks ($77.9^\circ$ after 35\,min from the beginning) and stays above $35^\circ$.

The full-sky image of the NCP simulations is shown in Fig.~\ref{fig:NCP_sim}, with a zoom-in of the NCP main field (top-right panel) and into the Cyg\,A field (bottom-right panel). Cyg\,A is $49.3^\circ$ away from the NCP, and, because of that, the primary beam attenuates its flux by a factor of ${\approx}10^{-4}$. However, its sidelobes are still strong and extend across the whole image, reaching the NCP main field and affecting its residuals if the source is not well modelled and subtracted. The same holds for Cas\,A, the effect of which could be even larger because it is closer to the NCP ($31.2^\circ$ away) and has a higher apparent brightness. 

\section{NCP data processing}\label{sec:ncp-processing}

Both data sets have been processed through an updated LOFAR-EoR production pipeline, which steps are extensively described by \citet{mertens_etal:2020} and Mertens et al.\ (in prep.). The pipeline consists essentially of (1) pre-processing and RFI removal, (2) initial calibration, (3) direction-dependent (DD) solving and DD sky-model subtraction, (4) imaging, (5) removal of residual foregrounds, and (6) power spectrum estimation. In this work, we skipped step (1) because we did not include any RFI into the simulated measurement set and we start with the already pre-processed observed data from \citet{mertens_etal:2020}. Step (5), which is performed with the Gaussian process regression (GPR) method \citep{mertens_etal:2018,mertens_etal:2024,hothi_etal:2021,acharya_etal:2024} was also skipped for both simulated and observed data sets, because this step does not use a sky model and we therefore do not need it to make the final comparison. In the following, we briefly describe the steps used for this work.

\subsection{Initial calibration}

The main goal of the initial calibration is to set the correct flux scale and correcting the sky-averaged phase offset per station. For the NCP, this is done in an unconventional way, by obtaining calibration solutions for two directions. One direction covers 3C\,61.1, which is the brightest source in the main field with a flux density of ${\approx}38\,\text{Jy}$ at 150\,MHz \citep{baldwin_etal:1985} and close to the first null of the primary beam, making it necessary to solve the varying DD effects of the primary beam and imperfections in the LOFAR primary beam model. The other direction includes the rest of the field, where only sources brighter than 35\,mJy are selected to reduce the computing time while keeping a high enough signal-to-noise ratio. This reduces the NCP sky model used for our real observations from 28\,773 to 1416 sources, thereby reducing the computational cost. The remaining sources account for only 1\% of the total flux model, and this has therefore only a small effect. For consistency, we performed a similar procedure for the sky model used for the simulations, but such a flux cut has a relatively small effect here, because the model has fewer sources (i.e.\ from 684 to 666 components). During the initial calibration, we only use baselines in the range $50{-}5000\lambda$. The lower limit was applied in real observations to avoid poor $uv$-coverage and the diffuse Galactic emission which is not present in our model and would bias the calibration \citep{patil_etal:2017}, while the upper limit minimises ionospheric effects and avoids resolving compact sources \citep{patil_etal:2016, mevius_etal:2022}.

The initial calibration was performed with \textsc{sagecal}, that uses the space alternating generalised expectation maximisation (SAGE) algorithm \citep{kazemi_etal:2011}. It distributes the processing across multiple nodes using a consensus optimisation \citep{boyd_etal:2011}. This method applies a frequency smoothness regularisation with a third-order Bernstein polynomial \citep{farouki_rajan:1988,yatawatta:2019}, using an alternating direction method of multipliers \citep[ADMM;][]{yatawatta:2015,yatawatta:2016}. The level of frequency smoothness for each iteration is determined by the regularisation parameter $\rho$, whose value depends on the apparent sky brightness for each source cluster and is therefore direction dependent  \citep{mevius_etal:2022}. More details about \textsc{sagecal} and its capabilities can be found in \citet{yatawatta:2015,yatawatta:2016,yatawatta:2018}, \citet{yatawatta:2017}, and \citet{gan_etal:2023}. This smooth gain calibration was performed with a frequency and time solution interval of 195\,kHz (i.e.\ one solution per SB) and 30\,s, respectively, to address relatively fast direction-independent ionospheric effects on the longer baselines. We note, though, that we did not include any ionospheric effects in our simulations. Recent studies have shown that the effect of the ionosphere is currently still expected to be below the thermal noise level \citep{vedantham_koopmans:2015,vedantham_koopmans:2016,brackenhoff_etal:2024}. The data were corrected using the solutions from the main field.

After this high-temporal and smooth-spectral resolution calibration, we performed a low-temporal and high-spectral resolution bandpass calibration with the same sky model to solve for the fine frequency response of the instrument. This was done using a single solution per SB over a time interval of 3\,h to mitigate the introduction of frequency structures due to an incomplete or inaccurate model as much as possible. We still used \textsc{sagecal}, but each SB was solved independently and no smoothness was applied to the solutions. As before, only the main field solutions were applied to the data.

After the bandpass calibration, all visibilities with amplitude higher than 70\,Jy were flagged to prevent \textsc{sagecal} from diverging from the correct solutions in the subsequent DD calibration step \citep{mertens_etal:2020}. We also averaged the data to a time resolution of 10\,s to speed up the next step, which is the most demanding step of the pipeline. 

\subsection{Direction-dependent solving and source subtraction}

The wide field of view of LOFAR introduces DD effects due to, for example, the time-varying primary beam and the ionospheric effects, that have to be taken into account to perform a good foreground subtraction. This is done by clustering the sky model components into a number of directions and ensuring a spectral smoothness using the consensus optimisation of \textsc{sagecal}. These  clusters typically are $1{-2}^\circ$ in size to reach a sufficient signal-to-noise with the sources inside the cluster and to minimise spatially-varying DD effects. For the simulations, the NCP main field was divided into 14 clusters (see Table~\ref{tab:sim-model}), while for real observations it currently consists of 108 clusters, because the sky model extends up to $15^\circ$ from the NCP. Two extra clusters were added for Cas\,A and Cyg\,A, because, similarly to the 3C\,61.1 case, they are so bright and so far away that they need separate solutions. No flux limit was applied to the sky model during this DD step. The gains were solved for all clusters simultaneously using a frequency interval of 195\,kHz. We set solution time intervals between 2.5 and 20\,min for the main field directions and an interval of 2.5\,min for Cas\,A and Cyg\,A to capture faster primary beam changes. All the sources in the sky models are included in this step.

To avoid over-fitting, a $uv$-cut at $250\lambda$ was introduced between baselines used for calibration and 21-cm signal extraction, where only baselines in the range $250{-}5000\lambda$ were used during the DD solving. This cut causes excess noise on the uncalibrated baselines, which is reduced by enforcing spectrally smooth solutions \citep{yatawatta:2016,barry_etal:2016}. Using this approach, it has been shown that there is no signal loss on the baselines of interest and very limited excess power \citep{mouri_sardarabadi_etal:2019, mevius_etal:2022}.

In contrast to the initial calibration, after the DD gain solving we did not correct the data with the solutions. Instead, we used the gains when subtracting the sky model from the visibilities, to obtain the residuals that were finally used for creating power spectra. 

In order to compare the performance of the new Cyg\,A model with the old one, we performed the DD calibration steps for both Cyg\,A models. In addition, we did a third run in which we modelled Cyg\,A as a point source with a spectral energy distribution following a third-order logarithmic polynomial with the coefficients given in \mksixteent. This extreme case will demonstrate whether having a spatially resolved model is important, even when the source is not resolved on the baselines used for calibration. In fact, calibration can still be affected by the source structure if the signal-to-noise ratio is high enough, as in the case of Cyg\,A. The results from the simulations (see Sect.~\ref{sec:results-sim}) that used the new model during the DD steps are only a benchmark for the calibration efficiency within the LOFAR-EoR pipeline, and do not say anything about how good our new model is. After the DD calibration, we considered only the first 6\,h of the observed data, in order to better compare the results with the simulated data set.

\subsection{Imaging and conversion to Kelvin}

After calibration and foreground subtraction, we gridded and imaged each SB of the residual visibilities using \textsc{wsclean}, making an $(l,m,\nu)$ image cube. During the gridding, each visibility received equal weight and we applied a Kaiser-Bessel anti-aliasing filter with a kernel extent of 15 $uv$-cells, an oversampling factor of 4096, and 32 $w$-layers. These parameters ensure that any error arising from the gridding process is confined significantly below the expected 21-cm signal \citep{offringa_etal:2019a,offringa_etal:2019b,mertens_etal:2020}.

We produced Stokes I and V images with a natural weighting scheme. Additionally, we generated even and odd time-step Stokes V images, the difference of which is used to estimate the thermal noise variance. We used a baseline range of $0{-}600\lambda$ during the imaging, setting a pixel scale of 0.5\,arcmin and an image size of $512\times512$ pixels, which results in a field of view of $4.3\times4.3^\circ$. After gridding, only the gridded $uv$-values between 50 and 250$\lambda$ were used for the power spectrum.

The image cube, which has units of Jy/beam, has to be converted into units of brightness temperature, which is Kelvin. This is done by Fourier transforming the images into a gridded $(u,v,\nu)$ visibility cube and then using the conversion technique detailed by \citet{offringa_etal:2019b}. Making the visibility cube, we applied a spatial Tukey function with a $4^\circ$ diameter to focus on the centre of the primary beam, which has a FWHM of ${\approx}4.1^\circ$ at 140\,MHz, while also mitigating its uncertainties farther from the centre. The gridded visibilities (in units of Kelvin) and the number of visibilities in each $(u,v,\nu)$-cell were used to estimate the power spectrum.

\subsection{Power spectrum estimation}

If $T({\bf x})$ is the brightness temperature of a signal at the physical coordinate $\mathbf{x}$, its power spectrum as a function of the wavenumber $\mathbf{k}$ (units of $h\,\text{cMpc}^{-1}$) is defined as
\begin{equation}\label{eq:PS}
    P(\mathbf{k}) = V | \tilde{T}(\mathbf{k}) |^2 \, ,
\end{equation}
where $V$ is the observed comoving volume and $\tilde{T}$ is the discrete Fourier transform of the brightness temperature. This 3D power spectrum is usually in units of $\text{K}^2\,h^{-3}\,\text{cMpc}^3$. The components of the wavenumber $\mathbf{k}$ perpendicular and along the line of sight are respectively defined as \citep{morales_hewitt:2004,vedantham_etal:2012}:
\begin{equation}\label{eq:kperp_kpar}
    \mathbf{k}_\perp = \frac{2\pi}{D_M(z)} \, \mathbf{u}\, ,\ 
    k_\| = \frac{2 \pi H_0 \nu_{21} E(z)}{ c (1+z)^2}\,\eta\, ,
\end{equation}
where $D_M(z)$ is the transverse comoving distance at redshift $z$, $\nu_{21}=1420\,\text{MHz}$, $H_0$ is the Hubble constant, $E(z)$ is the dimensionless Hubble parameter, $c$ is the speed of light, and $\eta$ is the Fourier dual of the frequency $\nu$ (often referred as delay).

The power $P(\mathbf{k})$ can be averaged in cylindrical shells of radius $k_\perp =|\mathbf{k}_\perp|$ to get the cylindrically-averaged (2D) power spectrum:
\begin{equation}\label{eq:PS2D}
    P(k_\perp, k_\|) = \frac{\sum_{\mathbf{k} \in (k_\perp,k_\|)} P(\mathbf{k})}{N_{(k_\perp,k_\|)}}\, ,
\end{equation}
where $N_{(k_\perp,k_\|)}$ is the number of $\mathbf{k}$ voxels falling into the $(k_\perp,k_\|)$-bin. If we choose spherical shells of radius $k=|\mathbf{k}|=\sqrt{k_\perp^2 + k_\|^2}$, a spherically-averaged (1D) power spectrum can also be obtained: 
\begin{equation}\label{eq:PS1D}
	\Delta^2(k) = \frac{k^3}{2\pi^2} \, \frac{\sum_{\mathbf{k} \in k} P(\mathbf{k})}{N_{k}}\, ,
\end{equation}
where $N_k$ is the number of $\mathbf{k}$ voxels falling into the $k$-bin. This is the power spectrum estimator used to set 21-cm upper limits because it is in units of $\text{mK}^2$. Both cylindrical and spherical power spectra are weighted by the gridded visibility thermal noise noise \citep[see][for more details]{mertens_etal:2020}.

Because $k_\|$ is essentially the Fourier dual of the frequency, a spectrally smooth signal is confined to low $k_\|$ values. This is the case for Galactic and extra-galactic foreground emission, which is expected to be dominated by synchrotron radiation at low frequencies and spectrally smooth over tens of MHz. On the other hand, the 21-cm signal is fluctuating rapidly in frequency, and its power extends over all $k_\|$ \citep[e.g.][]{santos_etal:2005}. However, every interferometer is intrinsically chromatic because a single $uv$-cell is crossed by different baselines at different frequencies, and this effect becomes stronger at longer baselines, corresponding to larger $k_\perp$ \citep{morales_etal:2012,morales_etal:2019}. Because of this, the foreground power is spread at higher $k_\|$, generating a characteristic shape known as the `foreground wedge' in the cylindrically-averaged power spectrum \citep{datta_etal:2010,liu_etal:2014a,liu_etal:2014b}. The maximum extent of the wedge in $k_\|$ is given by the horizon line, that can be accurately calculated in a non-flat sky formalism knowing the horizon delay \citep{munshi_etal:2025}. In the same way, we could find the delay lines for every direction in the sky \citep{munshi_etal:2025}. We then expect that all the sky emission lies in the wedge below the horizon line, leaving a foreground-free region called `EoR window'. However, the finite bandwidth of the data causes the Fourier transformation to leak some power above the wedge, requiring some frequency window function to reduce such a leakage \citep[e.g.][]{vedantham_etal:2012}. Also calibration errors might introduce rapid frequency fluctuations that contaminate the EoR window. This effect can be limited by using frequency smoothed gains, as we did in the first calibration step and during the DD solving.

In this work, we estimated the power spectra using the power spectrum pipeline \textsc{pspipe}\footnote{\url{https://gitlab.com/flomertens/pspipe}}. We used only gridded visibilities in the $50{-}250\lambda$ baseline range and applied a Hann window function to the frequency bandwidth. The reported power spectrum uncertainties were estimated from the sample variance, as described by \citet{mertens_etal:2020}.

\section{Results}\label{sec:results}

In this section, we show the results from both the simulated and observed data sets, comparing the residuals after the DD subtraction and their power spectra. We also compare results from the three Cyg\,A models described previously: the old, new, and point source models. All the cylindrical power spectra are shown as a ratio between the Stokes I and the thermal noise power spectrum, where the thermal noise was evaluated from time differences of Stokes V images. This ratio removes most of the effects that are caused by the sensitivity response of the instrument, allowing us to focus on the differences between the different Cyg\,A models. 

In the next section, we sketch how errors in the sky model could propagate into the foreground-subtracted visibilities after the DD subtraction. This is an important aspect to better understand the following results.

\subsection{DD gain errors due to model incompleteness}

In the following sections, the effect of model incompleteness for Cyg\,A, after its subtraction, is illustrated both in the image domain via residual images and in the power spectrum domain. The latter is particularly relevant for 21-cm cosmology experiments on baselines where the 21-cm signal is expected and illustrates that the effect in the image domain can be extremely subtle because continuum images contain all baselines and are averaged over frequency, thereby averaging out frequency-dependent fluctuations. In the cylindrical power spectra (Sects.~\ref{sec:results-sim-cps} and \ref{sec:results-obs-cps}), however, the effects as a function of baseline (proportional to $k_\perp$) and frequency (or its Fourier dual, `delay', proportional to $k_\parallel$) become much more evident.

In general, when determining LOFAR station-based gains, especially their DD gains, the incompleteness of the sky model can have a significant impact on the quality of the gains as a function of baselines and frequency \citep{patil_etal:2016,ewall-wice_etal:2017,mouri_sardarabadi_etal:2019,mevius_etal:2022}. When solving for multiple directions, the gains can modify the point spread function (PSF) used for subtraction in such a way that its sidelobes absorb unmodelled structures in the sky, thereby removing diffuse foregrounds and 21-cm emission on larger scales (see \citealt{patil_etal:2016} for a demonstration). This is why spectrally smooth gain regularisation was introduced \citep[e.g.][]{yatawatta:2015,gan_etal:2023}, as it suppresses the ability of gains to remove unmodelled source structures.

Similarly, very bright sources far from the phase centre (e.g.\ Cas\,A and Cyg\,A) can have a dominant impact via their PSF sidelobes, leaking power into the main target field (e.g.\ NCP or 3C\,196), as demonstrated by \citet{gan_etal:2022} and \citet{munshi_etal:2024:ul}. Hence, an incomplete sky model for these bright sources might lead to incorrect DD solutions in their direction. When subtracting them with their DD gains applied, the PSF-convolved source model will be incorrectly subtracted, leaving residuals in the main field of interest (NCP in our case). For fainter sources, this effect is much smaller and averages out when there are many sources in various directions and at different distances. However, Cyg\,A and Cas\,A are two extremely dominant sources, and their effect does not average away \citep{gan_etal:2022,munshi_etal:2024:ul}.

The errors on these source models can then enter the power spectrum in several different and complex ways: (i) even if the DD solutions in the direction of a bright source were known perfectly, applying them to an incorrect source model would leave residuals in the main field via the sidelobes, as discussed above; (ii) in practice, however, the incomplete sky model (not just the bright source model) also leads to complex gain errors, even in the case of strong gain regularisation as a function of frequency \citep[see e.g.][]{yatawatta:2015}. The latter effect can redistribute power across baselines depending on which baselines exhibit the largest differences between data and model, as well as on which baselines have the most visibilities (both contributing to the minimisation of the penalty function for the gain solutions).

In the following sections, we illustrate these effects for the case of an incomplete model for Cyg\,A.

\subsection{Results from simulations}\label{sec:results-sim}

\begin{figure*}[h!]
    \centering
    \includegraphics[width=\textwidth]{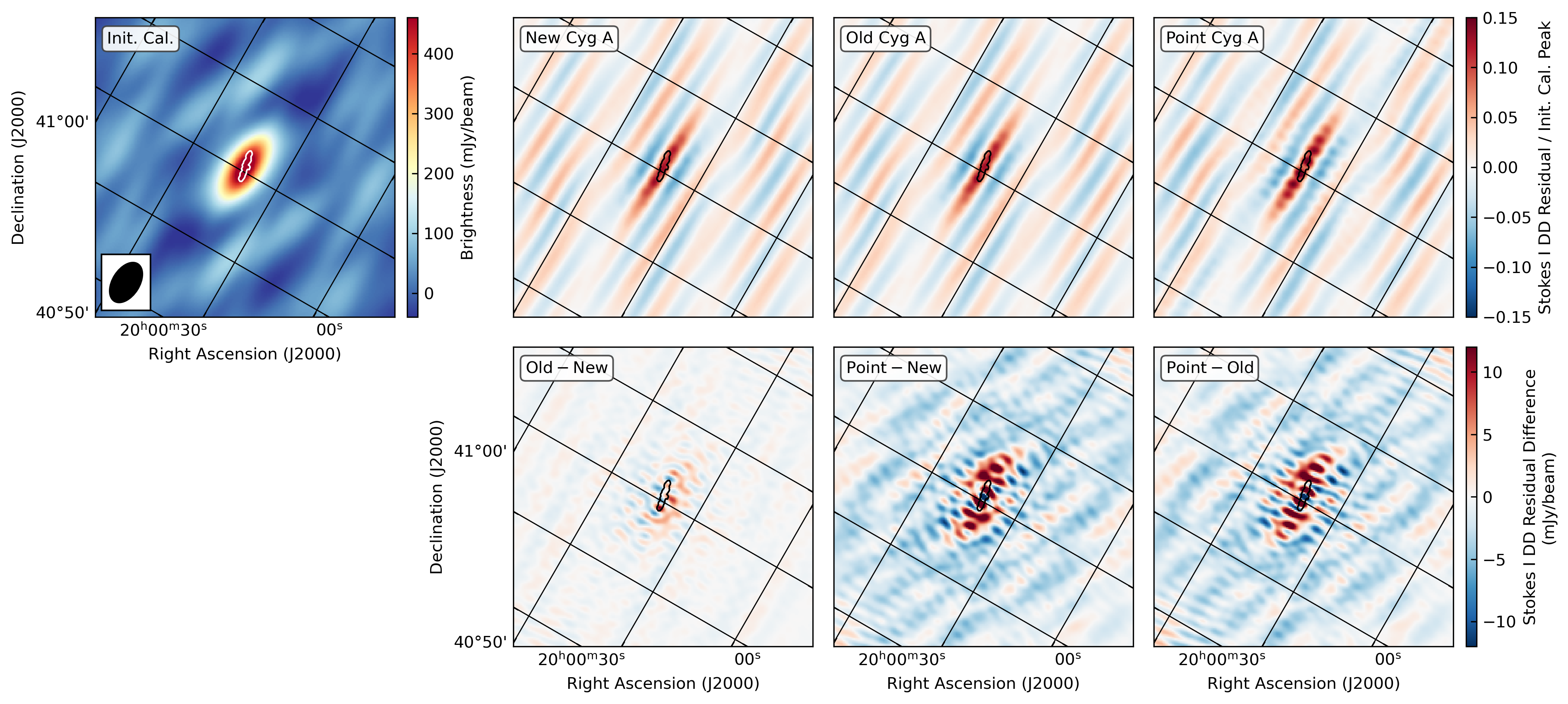}
    \caption{Dirty images of the simulated NCP data set in the direction of Cyg\,A. The top row shows, from left to right: (i) the image after initial calibration, and residual images after DD subtraction using the (ii) new, (iii) old and (iv) point source model, divided by the peak brightness of the initially calibrated image. The bottom row shows the differences between residual images obtained after DD subtraction for the three Cyg\,A models: $\text{Old}-\text{New}$ (left),  $\text{Point}-\text{New}$ (middle), and $\text{Point}-\text{Old}$ (right). All the images have been obtained with a natural weighting scheme and using a baseline range of $50{-}4500\lambda$, resulting in the synthesised beam shown in the bottom-left corner of the top-left panel. The contour in every panel is the 1\,Jy/beam level from Fig.~\ref{fig:CygA_imageHR_hotspots_peak} to show the position and extension of Cyg\,A.}
    \label{fig:CygA_residual_DDsub_sim}
\end{figure*}

After the DD subtraction step, we imaged the time and frequency averaged residuals at the Cyg\,A position in the simulated data set for all the three tested models. In the top row of Fig.~\ref{fig:CygA_residual_DDsub_sim}, from second to fourth panel, we show these residuals divided by the apparent peak brightness of Cyg\,A in the initially calibrated data, shown in the first panel. This gives an estimate of the relative errors using different models during the DD solving. In the second image, the model used for prediction and calibration was the same. The residuals therefore do not reflect the quality of the model, but the effectivity of calibration. The third and fourth panels show the extra residual power caused by (i) not modelling spectral indices, and (ii) assuming Cyg\,A is a point source. In the bottom row we also report the differences between the three residual images, to better highlight the structures affected by the different Cyg\,A models.

Although the old model is incomplete and spectrally incorrect, it leaves residual similar to the new model, both with a maximum ratio of approximately 0.13 close to the source centre. The point source model shows higher residuals, with a maximum of 0.17. The removal of all the three models generated ripples far from the sources and directed toward the NCP, eventually affecting the power spectrum. The intensity of these ripples are slightly larger for the extreme point source model, which also shows stronger small-scale fluctuations in the perpendicular direction. It is interesting to note that residuals for the new model are not fully consistent with the noise, even though the same model was used for both prediction and DD calibration. This highlights that even in the perfect scenario, calibration errors still affect the resulting residuals. This can be partly attributed to the added noise and the non-linear behaviour of calibration. However, in our experience, sources in the centre of the primary beam, are easier to subtract and leave lower residuals. A possible explanation for the relatively large residuals with a perfect model might be that the source passes through nulls in the primary beam and that not all stations have sufficient signal-to-noise at all times to accurately solve the source. 

\subsubsection{Cylindrical power spectra and ratios}\label{sec:results-sim-cps}

\begin{figure*}[h!]
    \centering
    \includegraphics[width=\textwidth]{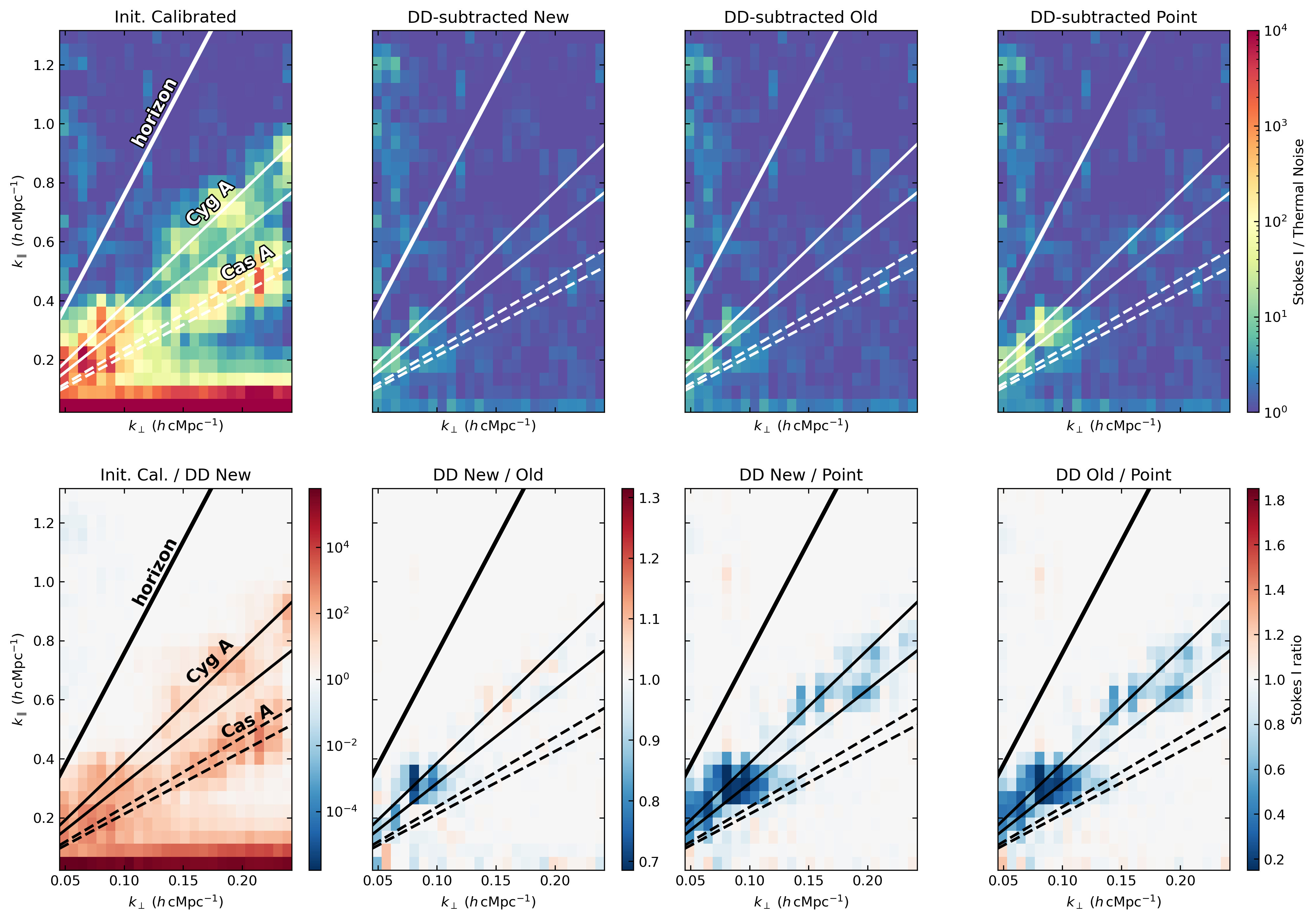} 
    \caption{Stokes I cylindrical power spectra divided by the thermal noise power spectrum at different stages of the processing of the simulated NCP data set. Top row, from left to right: power spectrum ratio after initial calibration, after DD subtraction using the new, old, and point source model of Cyg\,A. The bottom row shows ratios of results from processing with different models. From left to right: the initial calibration results over new model results, new model results over old model results, new model results over point source model results, and old model results over point source model results. `DD New / Point' and `DD Old / Point' share the same colour bar. In all the panels, the thick line indicates the horizon delay line (foreground wedge), whereas the thin solid and dashed line pairs indicate the delay ranges where we expect most of the power of Cyg\,A and Cas\,A, respectively.}
    \label{fig:PS2D_ALL_sim}
\end{figure*}

The cylindrical power spectra $P(k_\perp, k_\|)$, centred on the NCP, were estimated both after initial calibration and DD subtraction, and show that the DD steps removed most of the foreground emission within the wedge. The top row of Fig.~\ref{fig:PS2D_ALL_sim} shows the Stokes I power spectra divided by the thermal noise, estimated from time-differenced Stokes V. The thick line represents the horizon limit, above which the EoR window should be foreground-free. The other two pairs of lines, the solid and the dashed ones, show the range where we can maximally find Cyg\,A and Cas\,A, respectively. Even with error-free simulations, the residual power within the wedge is quite high after the DD subtraction, up to $10$ times the thermal noise. This might be due to calibration errors occurring during the DD solving or inaccuracies in the primary beam prediction, which also leaks power to higher $k_\|$, into the EoR window. Most of the residual power is along the Cyg\,A direction, while Cas\,A seems better subtracted. The maximum residual brightness of Cas\,A is 1\% of the peak of Cas\,A after the initial calibration and it is the same irrespective of which Cyg\,A model was used in the calibration. This shows that using spatially and spectrally different models of Cyg\,A during the DD calibration step does not considerably affect the gain calibration of other bright sources such as Cas\,A. The bottom-left panel of Fig.~\ref{fig:PS2D_ALL_sim} shows that the DD subtraction removes most of the power within the horizon. The residual power is usually taken into account with GPR, a step skipped in this work.

Ratios of DD subtracted power spectra for the three Cyg\,A models are shown in the bottom row of Fig.~\ref{fig:PS2D_ALL_sim}. Overall, the new model leaves the least residual power compared to the other two models. This is expected because the new model is the same model used during the prediction step. Most differences are visible around the Cyg\,A lines (thin solid lines), where the new model outperforms the old one, particularly at $k_\perp \lesssim 0.12\,h\,\text{cMpc}^{-1}$. The point source model, however, leaves the highest residuals across all $k_\perp$-$k_\|$ values within the Cyg\,A lines.

\begin{figure}
    \centering
    \includegraphics[width=1\columnwidth]{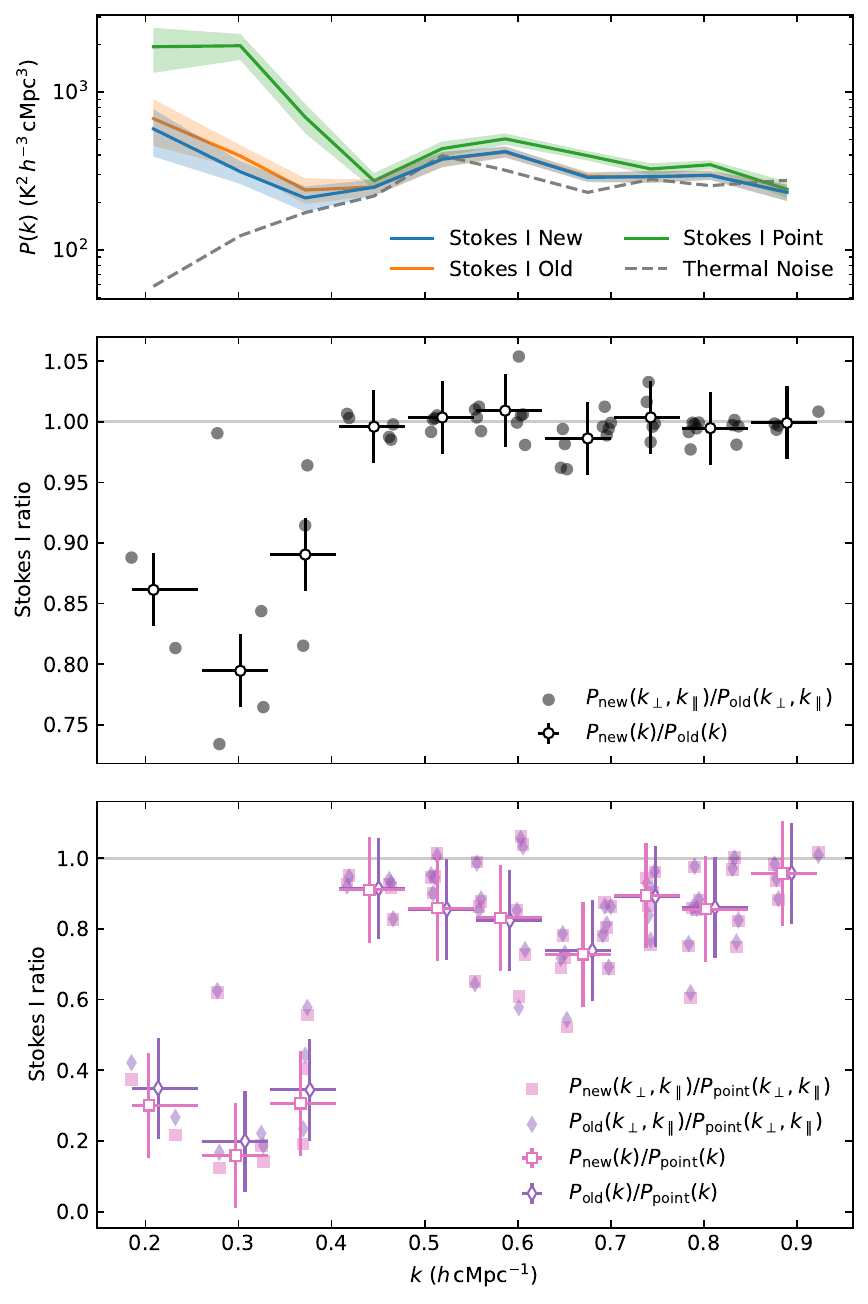}
    \caption{Power spectra along the Cyg\,A direction after DD subtraction (top panel) and their ratios (middle and bottom panels) for the simulated NCP data set. The top panel shows Stokes I $P(k)$ after the DD subtraction step using the new (blue), old (orange), and point source (green) models, along with the thermal noise level (dashed grey line). The shaded areas represent the $1\sigma$ uncertainties. The middle panel shows the ratio New/Old (black dots), while the bottom panel shows the ratios New/Point (pink squares), and Old/Point (purple diamonds). Filled markers represent the ratio of the cylindrical power spectra $P(k_\perp, k_\parallel)$ for each $(k_\perp, k_\parallel)$-cell within the Cyg\,A delay lines, while the white-faced markers indicate the ratios of the power spectra $P(k)$. For the latter, the horizontal error bars indicate the $k$-bin extension, while the vertical error bars indicate the $1\sigma$ standard deviation over all $(k_\perp, k_\|)$-cells of the ratios shown in Fig.~\ref{fig:PS2D_ALL_sim}. A small offset in $k$ has been added to the white-faced pink squares and purple diamonds in the bottom panel to avoid overlapping of error bars.}
    \label{fig:PS1D_DD_ratio_sim}
\end{figure}

The top panel of Fig.~\ref{fig:PS1D_DD_ratio_sim} shows the 1D power spectrum $P(k)$ for the three models, extracted from the cylindrical power spectra (i.e.\ top row of Fig.~\ref{fig:PS2D_ALL_sim}, but not divided by the thermal noise) along the Cyg\,A direction and averaged within $k$-bins of size $0.07\,h\,\text{cMpc}^{-1}$.\footnote{Our definition of $P(k)$ differs from $\Delta^2(k)$ due to how the averages are done. In fact, $P(k)$ is calculated as a simple average of $P(k_\perp, k_\|)$ without considering the weights, which are instead used to calculate $\Delta^2(k)$ from $P(\mathbf{k})$ (see Eq.~\ref{eq:PS1D}).} The middle and bottom panels show the ratios of these $P(k)$ pairs (white-faced markers), alongside the ratio values from bottom row of Fig.~\ref{fig:PS2D_ALL_sim} for each $(k_\perp,k_\|)$-cell within the Cyg\,A delay lines (filled markers). We plotted the ratio $P_\text{new}(k)/P_\text{old}(k)$ separately in the middle panel to make it easier to see where the new model has improved residuals compared to the old model. The $1\sigma$ uncertainties associated to $P(k)$ have been calculated as
\begin{equation}\label{eq:1Dps-err}
    P_\text{err}(k) = \sqrt{\frac{\sum_{(k_\perp,k_\|)\,\in\, k} {P_\text{err}^2(k_\perp,k_\|)}}{N_k^2}}\,, 
\end{equation}
where $N_k$ is the number of $(k_\perp,k_\|)$-cells falling in $k$-bin and $P_\text{err}(k_\perp,k_\|)$ is the error on the cylindrical power spectrum, estimated from the sample variance. For the ratio $P_i(k)/P_j(k)$, the sample variance is not a reliable uncertainty estimator, as all three data sets were drawn from the same sample and have the same noise. Therefore, the uncertainties of $P_i(k)/P_j(k)$ were set to the standard deviation over all $(k_\perp, k_\|)$-cells of the ratios shown in the bottom row of Fig.~\ref{fig:PS2D_ALL_sim}. The New/Old ratio reaches a minimum value of $P_\text{new}(k)/P_\text{old}(k)=0.80 \pm 0.03$ at $k=0.30\,h\,\text{cMpc}^{-1}$ and is significantly less than one for all $k<0.4\,h\,\text{cMpc}^{-1}$. At higher $k$ values, the ratio is close to one, suggesting that the new model does not offer improvement over the old model at smaller frequency and spatial scales. On the other hand, both $P_\text{new}(k)/P_\text{point}(k)$ and $P_\text{old}(k)/P_\text{point}(k)$ are consistently lower than one. The inverse-variance weighted averages across all $k$-bins are $\langle P_\text{new}(k) / P_\text{old}(k) \rangle = 0.95 \pm 0.05$, $\langle P_\text{new}(k) / P_\text{point}(k) \rangle \approx \langle P_\text{old}(k) / P_\text{point}(k) \rangle = 0.7 \pm 0.1$, indicating that the point source model performs worse than both the new and old models. This highlights that, although the point source model has correct spectral information, even a rough spatial characterisation of the source is necessary to improve DD calibration results.  

\subsubsection{Spherical power spectra and differences}\label{sec:results-sim-sps}

\begin{figure}
    \centering
    \includegraphics[width=1\columnwidth]{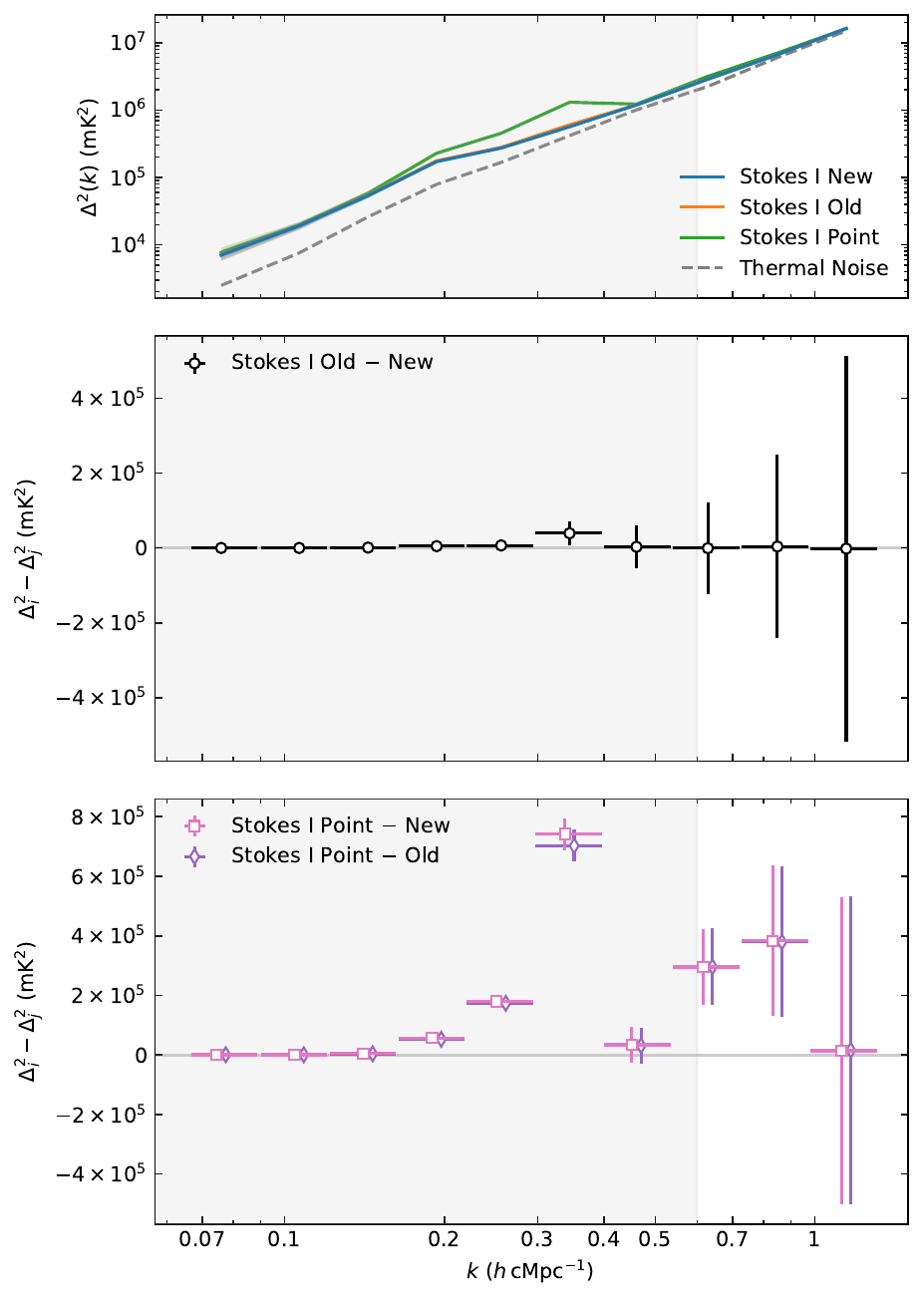}
    \caption{Spherical power spectra after DD subtraction (top panel) and their difference (middle and bottom panels) for the simulated NCP data set. The top panel shows Stokes I $\Delta^2(k)$ after the DD subtraction step using the new (blue), old (orange), and point source (green) Cyg\,A model, and the thermal noise level (dashed grey line). The shaded areas represent the $1\sigma$ uncertainties. The middle panel shows the difference $\Delta^2_\text{old} - \Delta^2_\text{new}$ (black dots), while the bottom panel shows the differences $\Delta^2_\text{point} - \Delta^2_\text{new}$ (pink squares), and $\Delta^2_\text{point} - \Delta^2_\text{old}$ (purple diamonds), with the associated $1\sigma$ uncertainties (values reported in Table~\ref{tab:merged-diff-results}). The horizontal error bars indicate the $k$-bin extension. We added a small offset in $k$ to the pink and purple markers in the bottom panel to avoid overlapping of error-bars. The grey shaded area delimits the range in $k$ where $\Delta^2$ upper limits are usually extracted from LOFAR data, being the most sensitive range for LOFAR.}
    \label{fig:PS3D_allDD_diff_sim}
\end{figure}

Although the power spectra $P(k)$ along the Cyg\,A direction give a robust estimation of the impact that the three different models have on the residual power, we saw in the third and fourth bottom panels of Fig.~\ref{fig:PS2D_ALL_sim} that differences are present also in different regions of the cylindrical power spectrum. We can then use Eq.~\eqref{eq:PS1D} to estimate the spherical power spectrum $\Delta^2(k)$ in $k$-bins spanning the whole $(k_\perp,k_\|)$-space and assess the overall impact of the three models on the 21-cm upper limits. 

The total Stokes I spherical power spectra are shown in the top panel of Fig.~\ref{fig:PS3D_allDD_diff_sim}, evaluated within ten logarithmically spaced $k$-bins in the range $0.08\le k \le 1.15\,h\,\text{cMpc}^{-1}$, with a bin size of $\text{d} k/k \approx 0.3$. The differences between model pairs $i$-$j$ and the associated $1\sigma$ uncertainties have been calculated as
\begin{equation}\label{eq:diff_ps3d}
    D_{ij} = \Delta_{i}^2 - \Delta_{j}^2\, ,\ D_{ij, \text{err}} = \sqrt{{(\Delta_{i, \text{err}}^2)}^2 + {(\Delta_{j, \text{err}}^2)}^2}\, ,
\end{equation}
where $\Delta^2_\text{err}(k)$ is estimated from the sample variance of the power spectrum (see $P_{\text{err}}$ in Eq.~\ref{eq:1Dps-err}). In this case, the sample variance is used also for the uncertainties of the differences, because $D_{ij}$ is calculated directly from $\Delta^2(k)$. The values are reported in Table~\ref{tab:merged-diff-results}, with also the inverse-variance weighted mean of the differences calculated over two representative $k$-bin ranges. The difference $\Delta_\text{old}^2 - \Delta_\text{new}^2$ (middle panel of Fig.~\ref{fig:PS3D_allDD_diff_sim}) is larger than zero across most of the $k$-bins, with just a few exceptions. However, all these values are consistent with zero when the sample variance error is taken into account, except at $k=0.35\,h\,\text{cMpc}^{-1}$ where $\Delta_\text{old}^2> \Delta_\text{new}^2$ significantly. We observe the strongest differences at this $k$-bin for all the model pairs, consistent with Fig.~\ref{fig:PS1D_DD_ratio_sim}. Looking at the differences with the point source model (bottom panel of Fig.~\ref{fig:PS3D_allDD_diff_sim}), we find that $\Delta_\text{point}^2$ is higher than $\Delta_\text{new}^2$ and $\Delta_\text{old}^2$ across all $k$, with significant values at $0.19\le k \le 0.85\,h\,\text{cMpc}^{-1}$.

\subsection{Results from real observations}\label{sec:results-obs}

\begin{figure*}
    \centering
    \includegraphics[width=1\textwidth]{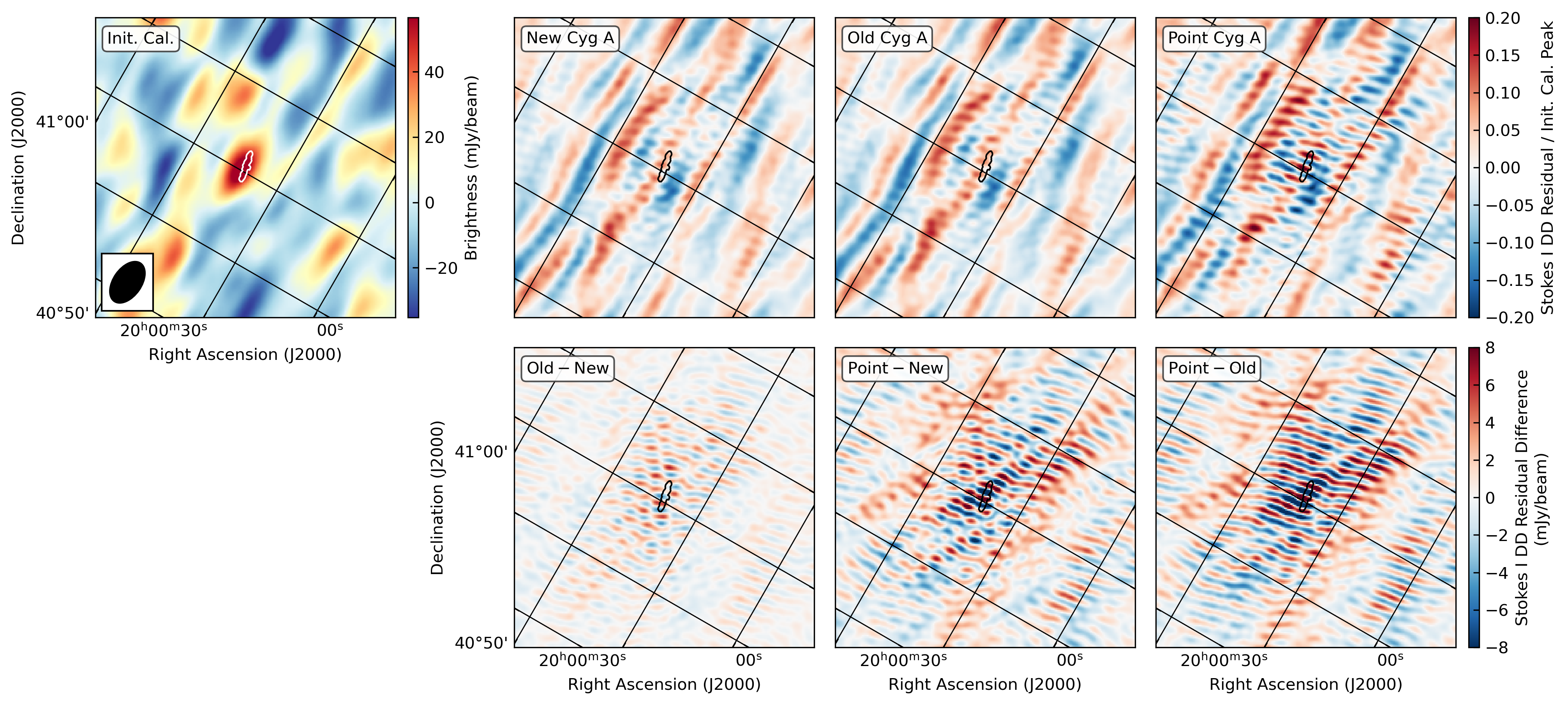}
    \caption{Dirty images of the real observed NCP data set in the direction of Cyg\,A. The top row shows, from left to right: (i) the image after initial calibration, and residual images after DD subtraction using the (ii) new, (iii) old and (iv) point source model, divided by the peak brightness of the initially calibrated image. The bottom row shows the differences between residual images obtained after DD subtraction for the three Cyg\,A models: $\text{Old}-\text{New}$ (left),  $\text{Point}-\text{New}$ (middle), and $\text{Point}-\text{Old}$ (right). All the images have been obtained with a natural weighting scheme and using a baseline range of $50{-}4500\lambda$, resulting in the synthesised beam shown in the bottom-left corner of the top-left panel. The contour in every panel is the 1\,Jy/beam level from Fig.~\ref{fig:CygA_imageHR_hotspots_peak} to show the position and extension of Cyg\,A.} 
    \label{fig:CygA_residual_DDsub_obs}
\end{figure*}

For the actual observations, we follow a similar route compared to the simulated data. In the top row of Fig.~\ref{fig:CygA_residual_DDsub_obs}, from second to fourth panel, we show the residuals at the Cyg\,A position for all the three tested models, divided by the apparent peak brightness of the initial calibrated Cyg\,A, shown in the first panel. Differences between the residual images are shown in the bottom row. The residuals left from the new and old model are similar, as we saw in the simulations, with a maximum ratio within the Cyg\,A region of 0.09 and 0.11, respectively. The point source model shows the highest residuals, with a maximum ratio of 0.23. Although the measured residuals for the new and old models are smaller than those in the simulations, the fluctuations originated from the subtraction of these models are stronger even far from the source and more complex. The subtraction of a single point component leaves the strongest sidelobes and ripples, both close and far from the source location. 

\subsubsection{Cylindrical power spectra}\label{sec:results-obs-cps}

\begin{figure*}
    \centering
    \includegraphics[width=1\textwidth]{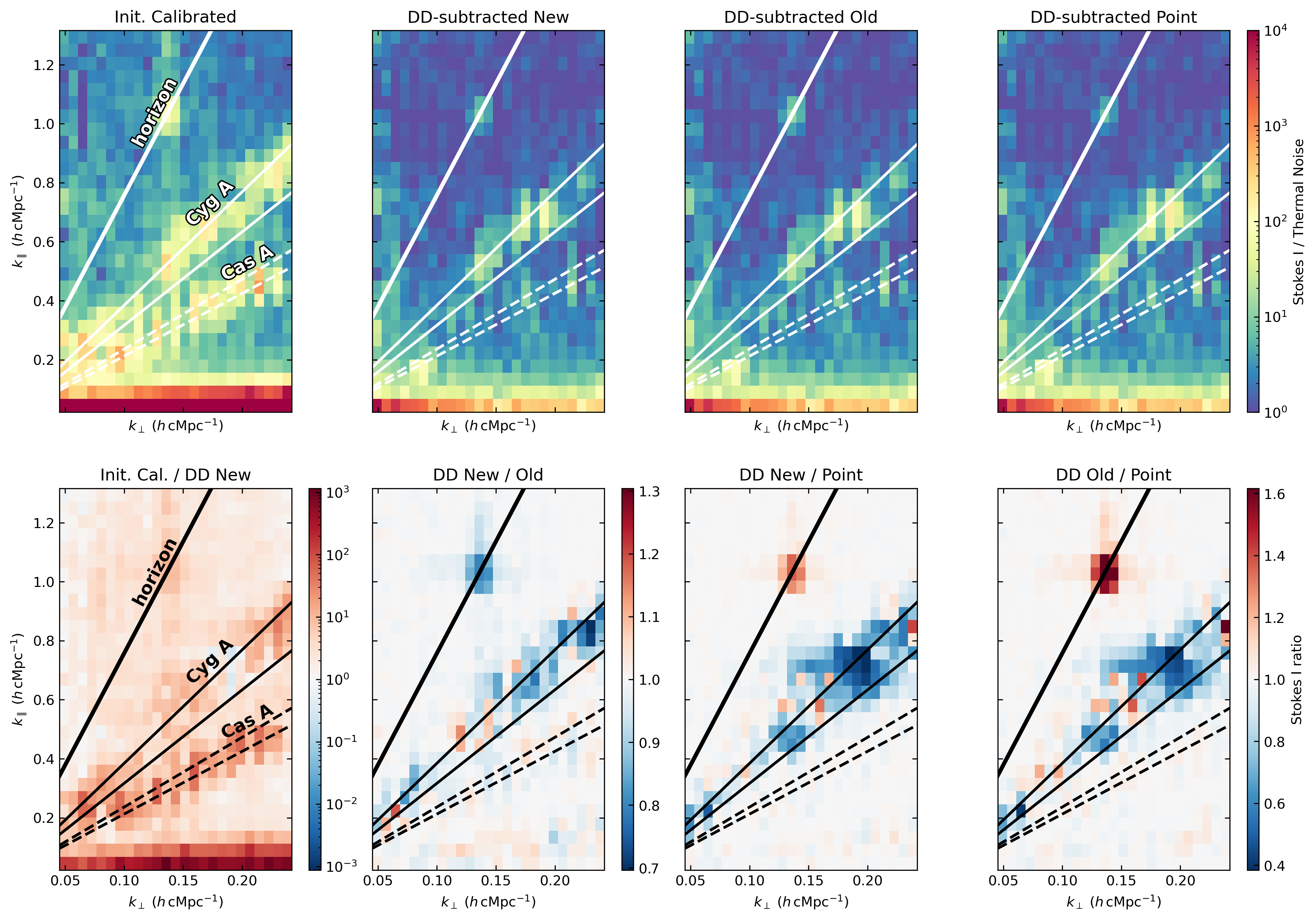}
    \caption{Stokes I cylindrical power spectra divided by the thermal noise power spectrum at different stages of the processing of the real observed NCP data set. Top row, from left to right: power spectrum ratio after initial calibration, after DD subtraction using the new, old, and point source model of Cyg\,A. The bottom row shows ratios of results from processing with different models. From left to right: the initial calibration results over new model results, new model results over old model results, new model results over point source model results, and old model results over point source model results. `DD New / Point' and `DD Old / Point' share the same colour bar. In all the panels, the thick line indicates the horizon delay line (foreground wedge), whereas the thin solid and dashed line pairs indicate the delay ranges where we expect most of the power of Cyg\,A and Cas\,A, respectively.}
    \label{fig:PS2D_ALL_obs}
\end{figure*}

The top row of Fig.~\ref{fig:PS2D_ALL_obs} shows the Stokes I cylindrical power spectra after the initial calibration and the DD subtraction, all divided by the thermal noise power spectrum. In this real case scenario, the DD subtraction is not able to remove some foreground emission at low $k_\|$. The power spectrum reaches $10^4$ times the thermal noise level, because of systematics such as calibration errors and sky model incompleteness, including diffuse Galactic emission not present in the model. However, the decrease in power after the DD subtraction step is a factor of $10^3$ at low $k_\|$. Some residual power is also observed along the Cas\,A direction, which might be the cause of some of the ripples visible in Fig.~\ref{fig:CygA_residual_DDsub_obs}.

In the bottom row of Fig.~\ref{fig:PS2D_ALL_obs} (second to fourth panel), we show the ratios of the DD subtracted power spectra for the three Cyg\,A models. In all three cases, most differences are confined along the Cyg\,A direction. Although residuals from Cas\,A are observed in the top panels, the three different Cyg\,A models do not affect them differently, and the power along the Cas\,A direction cancels out in the ratio. 

\begin{figure}
    \centering
    \includegraphics[width=1\columnwidth]{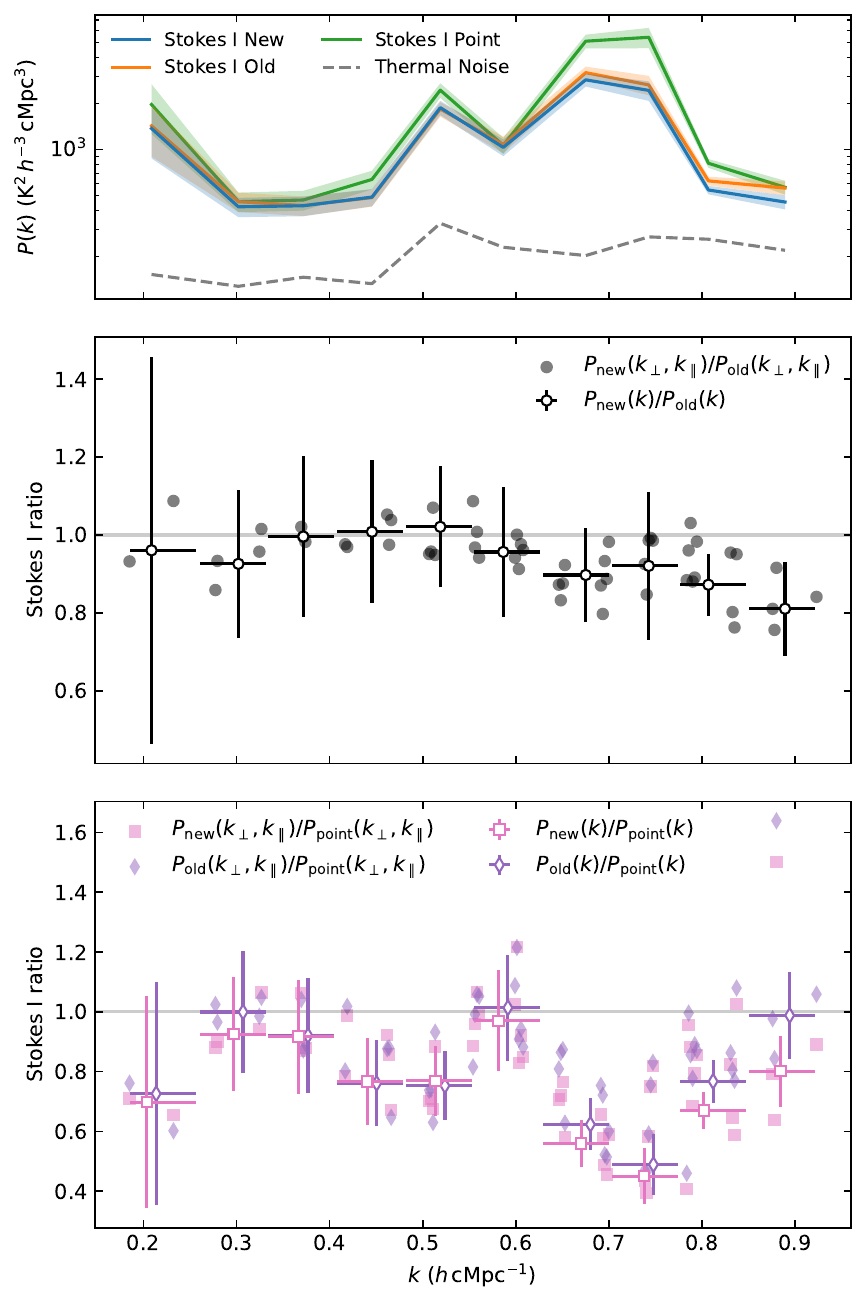}
    \caption{Power spectra along the Cyg\,A direction after DD subtraction (top panel) and their ratios (middle and bottom panels) for the real observed NCP data set. The top panel shows Stokes I $P(k)$ after the DD subtraction step using the new (blue), old (orange), and point source (green) models, along with the thermal noise level (dashed grey line). The shaded areas represent the $1\sigma$ uncertainties. The middle panel shows the ratio New/Old (black points), while the bottom panel shows the ratios New/Point (pink squares), and Old/Point (purple diamonds). Filled markers represent the ratio of the cylindrical power spectra $P(k_\perp, k_\parallel)$ for each $(k_\perp, k_\parallel)$-cell within the Cyg\,A delay lines, while the white-faced markers indicate the ratios of the power spectra $P(k)$. For the latter, the horizontal error bars indicate the $k$-bin extension, while the vertical error bars indicate the $1\sigma$ uncertainties of the ratio. A small offset in $k$ has been added to the white-faced pink squares and purple diamonds in the bottom panel to avoid overlapping of error bars.}
    \label{fig:PS1D_DD_ratio_obs}
\end{figure}

The 1D power spectra $P(k)$ along the Cyg\,A direction for the three models are shown in the top panel of Fig.~\ref{fig:PS1D_DD_ratio_obs}. The ratio $P_\text{new}(k)/P_\text{old}(k)$ is shown in the middle panel, while $P_\text{new}(k)/P_\text{point}(k)$ and $P_\text{old}(k)/P_\text{point}(k)$ are shown in the bottom panel. In this real case, the $1\sigma$ uncertainties associated to ratio $P_i(k)/P_j(k)$ have been estimated from the sample variance as
\begin{equation}
    \left[\frac{P_i}{P_j}\right]_\text{err} = \frac{P_i}{P_j}\sqrt{\left( \frac{P_{i,\text{err}}}{P_{i}}\right)^2 + \left(\frac{P_{j,\text{err}}}{P_j} \right)^2 }\,,
\end{equation}
where $P_\text{err}$ is from Eq.~\eqref{eq:1Dps-err}. In real observations, the new model results in a residual power spectrum similar to the old model at low $k$. The ratio $P_\text{new}(k)/P_\text{old}(k)$ reaches a minimum of $0.8 \pm 0.1$ at $k=0.89\,h\,\text{cMpc}^{-1}$, with an inverse-variance weighted average of $0.91 \pm 0.05$ over the entire $k$ range. In contrast, the point source model shows higher power at all scales, with $\langle P_\text{new}(k)/P_\text{point}(k) \rangle = 0.67 \pm 0.03$ and $\langle P_\text{old}(k)/P_\text{point}(k) \rangle = 0.74 \pm 0.04$. These values are consistent with those measured from the simulations. Interestingly, the new model leaves lower residuals than the old one at high $k$, whereas in the simulations we observed the opposite trend, with larger differences at small $k$. This discrepancy results from the DD calibration, which successfully removed most of the Cyg\,A power at $k_\perp < 0.12\,h\,\text{cMpc}^{-1}$ in the observed data set (see Fig.~\ref{fig:PS2D_ALL_obs}), but not in the simulated one, where the cylindrical power spectra where noise-like at higher $k_\perp$ (see Fig.~\ref{fig:PS2D_ALL_sim}).

\subsubsection{Spherical power spectra and differences}\label{sec:results-obs-sps}

\begin{figure}
    \centering
    \includegraphics[width=1\columnwidth]{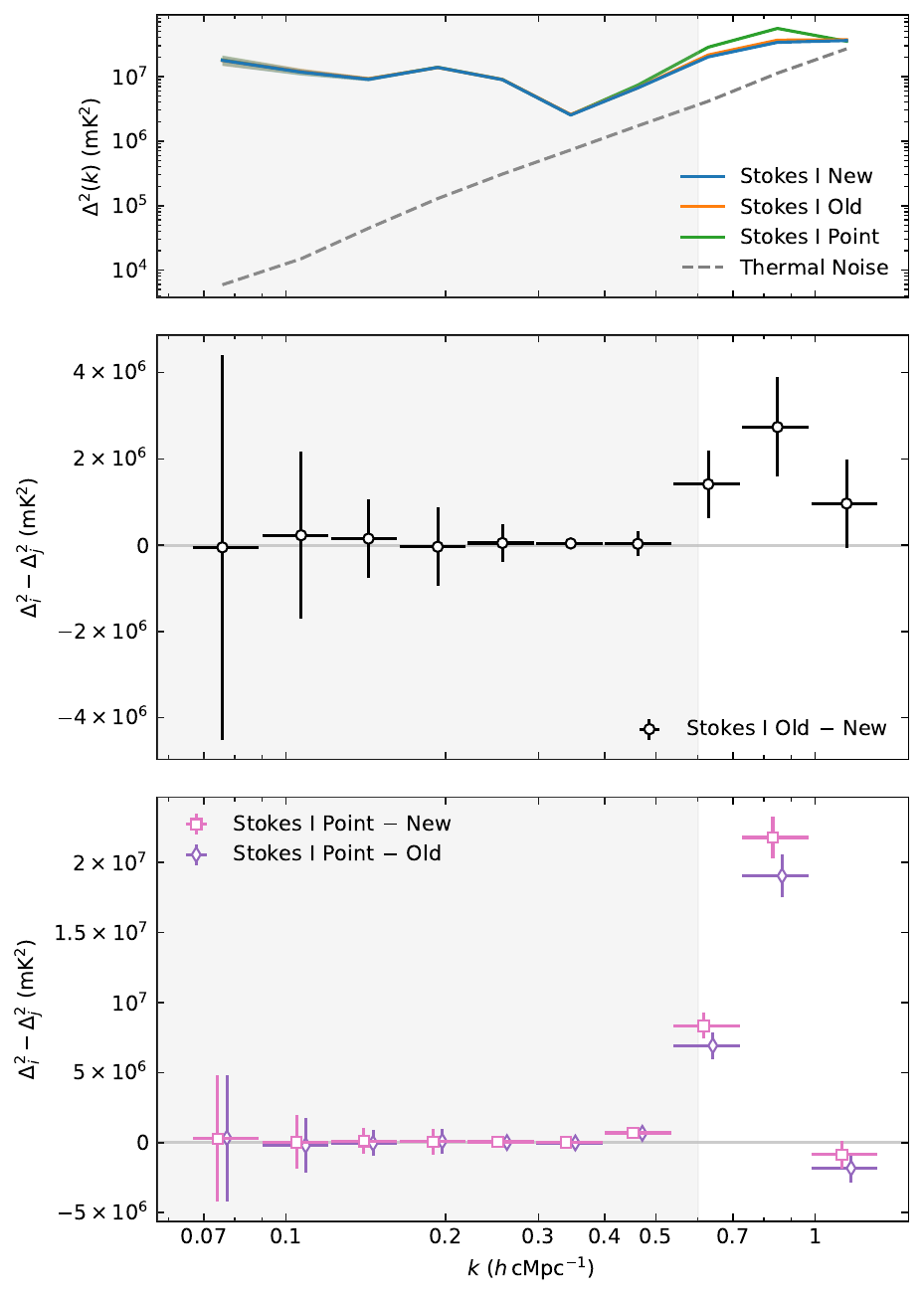}
    \caption{Spherical power spectra after DD subtraction (top panel) and their difference (middle and bottom panels) for the real observed NCP data set. The top panel shows Stokes I $\Delta^2(k)$ after the DD subtraction step using the new (blue), old (orange), and point source (green) Cyg\,A model, and the thermal noise level (dashed grey line). The shaded areas represent the $1\sigma$ uncertainties. The middle panel shows the difference $\Delta^2_\text{old} - \Delta^2_\text{new}$ (black dots), while the bottom panel shows the differences $\Delta^2_\text{point} - \Delta^2_\text{new}$ (pink squares), and $\Delta^2_\text{point} - \Delta^2_\text{old}$ (purple diamonds), with the associated $1\sigma$ uncertainties (values reported in Table~\ref{tab:merged-diff-results}). The horizontal error bars indicate the $k$-bin extension. We added a small offset in $k$ to the pink and purple markers in the bottom panel to avoid overlapping of error-bars. The grey shaded area delimits the range in $k$ where $\Delta^2$ upper limits are usually extracted from LOFAR data, being the most sensitive range for LOFAR.}
    \label{fig:PS3D_allDD_diff_obs}
\end{figure}

The total Stokes I spherical power spectra $\Delta^2(k)$ are shown in the top panel of Fig.~\ref{fig:PS3D_allDD_diff_obs}. As already seen in the top panel of Fig.~\ref{fig:PS2D_ALL_obs}, the power is a few orders of magnitude higher than the thermal noise at $k<0.3\,h\,\text{cMpc}^{-1}$, which corresponds to approximately $k_\|< 0.3\,h\,\text{cMpc}^{-1}$ in the cylindrical power spectrum. Here, the foreground emission within and close to the main lobe of the primary beam has not been completely removed, either because of calibration errors or incomplete sky model (e.g.\ extended Galactic emission is not removed and may affect low $k_\perp$). At higher $k$, $\Delta^2$ is closer to the thermal noise, even if some excess power is present because of poorly subtracted off-axis sources, mainly Cyg\,A and Cas\,A. In fact, the difference between model pairs, shown in the middle and bottom panels of Fig.~\ref{fig:PS3D_allDD_diff_obs} and evaluated with Eq.~\eqref{eq:diff_ps3d}, increases in these $k$-bins, reaching peaks of ${\approx}(1600\,\text{mK})^2$ for $\Delta_\text{old}^2 - \Delta_\text{new}^2$ and ${\approx}(4500\,\text{mK})^2$ for both $\Delta_\text{point}^2 - \Delta_\text{new}^2$ and $\Delta_\text{point}^2 - \Delta_\text{old}^2$ at $k=0.85\,h\,\text{cMpc}^{-1}$. The real observed data do not prefer one model at $k\le 0.35\,h\,\text{cMpc}^{-1}$, where all the differences have large uncertainties and are consistent with zero. This is in contrast with the simulations, where differences and uncertainties were smaller. This happens because in the observed data there are other systematics, such as model incompleteness, calibration errors, and inaccuracies in the primary beam modelling, that play a bigger role during the DD subtraction. Given the current situation of the systematics, we do not expect to see larger differences by using different Cyg\,A models. However, an overall improvement is achievable by using the new model instead of the old one in the range $0.63{-}1.15\,h\,\text{cMpc}^{-1}$, where $\langle \Delta_\text{old}^2 - \Delta_\text{new}^2 \rangle= (1259 \pm 742\,\text{mK})^2$. At lower $k$, the average difference is consistent with zero, but the expected value is ${\approx} (257\,\text{mK})^2$ in the range $0.08{-}0.63\,h\,\text{cMpc}^{-1}$. This shows that, in principle, the new Cyg\,A model is capable of reducing the excess power by hundreds of mK, even if $\Delta^2_\text{new}$ is still 2--3 times higher than the thermal noise level.

\section{Discussion and conclusions}\label{sec:conclusions}

The latest upper limits on the 21-cm power spectrum from the LOFAR-EoR KSP show an excess power in $\Delta^2(k)$ with respect to thermal noise \citep{mertens_etal:2020}. \citet{gan_etal:2022} analysed various causes of such an excess, concluding that it might be dominated by residuals of bright sources, especially those in the far field. Their power spectra, even after GPR foreground removal, show non-negligible power in the Cas\,A and Cyg\,A direction, hinting that the modelling and subtraction of these A-team sources has been performed poorly. The models of these sources, also used by the current LOFAR-EoR processing pipeline, are low resolution, with no spectral variation and incorrect spectral indices. These models have been obtained from VLA data at ${\approx} 74$\,MHz and assigned a fixed spectral index of $-0.8$ to each component, which is used to extrapolate values to the LOFAR HBA band where the 21-cm power spectrum is calculated. 

In this work, we focused on Cyg\,A, the brightest of the A-team at 150\,MHz, and used it as a test case to asses the impact of a better spatial and physical model on the 21-cm power spectrum. In Sect.~\ref{sec:cyga-model}, we introduced what we called the `old' Cyg\,A model and showed that extrapolating it to the HBA frequency range results in 16\% missing flux. Not including any spectral variation across the source is incorrect, because we know from \mksixteent\ that the brightest hotspots (A and D) show a frequency turnover at around 140--150\,MHz. To solve these issues, we made a new model directly from HBA observations (110--250\,MHz), taking advantage of the high spatial resolution allowed by the Dutch LOFAR configuration, i.e.\ ${\approx}5$\,arcsec, and using a second-order logarithmic polynomial to model the spectral behaviour of each component of the source (Sect.~\ref{sec:new-model}). We used the forced-spectrum method of \textsc{wsclean} to transfer this spectral information into the model during a multi-scale and multi-frequency deconvolution \citep{ceccotti_etal:2023}. This makes our new Cyg\,A model the first test case of such a method for improving 21-cm power spectrum results. 

In Sect.~\ref{sec:results}, we compared the performances of the old model with our new one within the LOFAR-EoR processing pipeline (Sect.~\ref{sec:ncp-processing}), using both simulated and observed data of the NCP field. Additionally, we considered an extreme Cyg\,A model, constituted by a single point source at the Cyg\,A location but with correct spectral information. On the simulated data, the most significant differences appear in $P(k)$ along the Cyg\,A direction at $k<0.4\,h\,\text{cMpc}^{-1}$, where the new model performs better than both the old and point source models. These differences are washed out in $\Delta^2(k)$, where all $(k_\perp, k_\|)$-cells are considered, resulting in $\Delta_\text{new} \approx \Delta_\text{old}$ across most $k$ bins. For the real observed data, differences between the three models in both $P(k)$ along the Cyg\,A direction and $\Delta^2(k)$ are less pronounced at $k<0.4\,h\,\text{cMpc}^{-1}$, where uncertainties make their power spectra consistent with one another. Larger differences are observed at higher $k$, where the point source model results in significantly higher power in both the power spectrum estimators. The lowest residuals are achieved with the new model at $k>0.6\,h\,\text{cMpc}^{-1}$, where the old and point source models show an average excess of more than $(1250\,\text{mK})^2$ and $(2300\,\text{mK})^2$, respectively.

As we did not inject any systematics in the simulations, apart from adding the thermal noise, we expect all the differences to come from the different Cyg\,A models. In the real observations, where other systematics are present, these systematics have a larger impact on the residuals, hiding the effect of the different models, especially at the lower $k$ where upper limits on the 21-cm power spectrum are usually extracted. We know that Cyg\,A is the strongest source at 150\,MHz, but not in apparent flux for NCP observations. There are many other bright sources outside the main lobe of the primary beam that could leave subtraction residuals strong enough to outweigh those left from Cyg\,A, most notably Cas\,A, which has an apparent flux approximately three times higher than Cyg\,A. The DD subtracted plots in Fig.~\ref{fig:PS2D_ALL_obs} show that some residuals associated with Cas\,A are present, highlighting the need of a better model for this source as well. However, as long as we focus our analysis at $k<0.6\,h\,\text{cMpc}^{-1}$, as in the analysis of real data, we do not expect significant improvements until the effects of other systematics are lowered. 

In conclusion, we have demonstrated that the forced-spectrum method can generate spectrally accurate models of complex sources such as Cyg\,A, provided that physical spectral index maps are available or can be extracted from the data. This approach ensures that robust spectral information is embedded in a high-resolution model, which is fundamental for using such a model at different frequencies from those at which the model was originally extracted. For example, we showed that while the old model has a correct flux density at 74\,MHz, its incorrect spectral index results in a much lower extrapolated flux density at 150\,MHz than the expected value. Moreover, the new model is described as a standard sky model, with a low number of point and Gaussian components with spectral information, and can therefore directly be used in calibration. The new Cyg\,A model presented in this work, and more generally, the forced-spectrum fitting method, could greatly benefit high-resolution and wide-field interferometers operating at low frequencies, such as the upcoming SKA-Low. This advancement will be beneficial for enhancing the accuracy and reliability of astronomical observations at low frequencies.

Because no GPR analysis was applied before power spectrum estimation, all the improvements that we measured in the 21-cm power spectrum come from the DD calibration, the only step where we used Cyg\,A. Alongside with assessing the impact of a better model on the power spectrum, we demonstrated that the calibration itself can benefit from a forced-spectrum model. We cannot make a one-to-one comparison with \citet{mertens_etal:2020} about the excess power in the LOFAR power spectrum, because we did not apply a GPR technique. However, we can conclude that improving the modelling of Cyg\,A can reduce the excess power up to $(250\,\text{mK})^2$ over the $k=0.08{-}0.63\,h\,\text{cMpc}^{-1}$ range, as measured in the real observations (see Table~\ref{tab:merged-diff-results}). This reduction can be smaller after GPR, because some of the extra power that we observed would be absorbed in the foreground mode-mixing kernel, making the difference between the three models less strong. Our simulation scenario can be expected to be comparable to the results of a GPR analysis, because we have not included the systematics that GPR removes. Here, indeed, the measured differences are smaller, with $\langle\Delta_\text{old}^2 - \Delta_\text{new}^2 \rangle \approx (17\,\text{mK})^2$ over the $k=0.08{-}0.63\,h\,\text{cMpc}^{-1}$ range. What we showed is that using both spatially and spectrally accurate models of bright off-axis sources is important to lower the upper limits on the 21-cm power spectrum. 

In order to see a significant impact of a better spectral modelling in real data, other systematics must be reduced first. Because we flagged data after the DD subtraction, we expect even low-level RFI to be removed and they would therefore not be a dominant source of excess power. Another possible cause of the excess power might be ionospheric effects that have not been completely solved for in the initial calibration step. Although, \citet{gan_etal:2022} and \citet{brackenhoff_etal:2024} concluded that the ionosphere is currently not the dominant source of the excess power. In the analysis of \citet{brackenhoff_etal:2024}, it is shown that rapid spatial and temporal variations in the primary beam model could, however, result in incorrect DD calibration solutions. This is  especially true for directions far from the field centre, where primary beam sidelobes move faster in time and have steep edges with deep nulls. Whereas we do not see incorrect DD gains, we see bright residuals associated with Cyg\,A even in the simulations where the calibration model is the same as the one used in the prediction step (first panel of Fig.~\ref{fig:CygA_residual_DDsub_sim}). This could mean that the DD calibration gains are not able to capture the finest spatial, temporal, and spectral variations of the source due to the primary beam sidelobes. Understanding and characterising the time and frequency behaviour of the LOFAR HBA primary beam appears to be the next important step in trying to reduce the excess power in the NCP field. Processing and analysing a different sky region, such as the 3C\,196 field, could be helpful as well in understanding the source of the systematics (Ceccotti et al.\ in prep.). 

\section*{Data availability}

The component list of the new model of Cygnus\,A is available at \url{https://doi.org/10.5281/zenodo.14863290}. The observed and simulated data underlying this article will be shared on reasonable request to the corresponding author.

\begin{acknowledgements}
EC, ARO and LVEK would like to acknowledge support from the Centre for Data Science and Systems Complexity (DSSC), Faculty of Science and Engineering at the University of Groningen. EC acknowledges support from the Ministry of Universities and Research (MUR) through the PRIN project `Optimal inference from radio images of the epoch of reionization'. SAB, SG, LVEK and SM acknowledge the financial support from the European Research Council (ERC) under the European Union's Horizon 2020 research and innovation programme (Grant agreement No.\ 884760, `CoDEX'). FGM acknowledges support from a PSL Fellowship. RG acknowledges support from SERB, DST Ramanujan Fellowship no.\ RJF/2022/000141. The post-doctoral contract of IH was funded by Sorbonne Université in the framework of the Initiative Physique des Infinis (IDEX SUPER). LOFAR, the Low Frequency Array designed and constructed by ASTRON, has facilities in several countries, that are owned by various parties (each with their own funding sources), and that are collectively operated by the International LOFAR Telescope (ILT) foundation under a joint scientific policy. In this work, we made use of the \textsc{kvis} \citep{gooch:1996} and \textsc{ds9} \citep{joye_mandel:2003} FITS file image viewers, and the \textsc{astropy} \citep{astropy_coll:2022}, \textsc{matplotlib} \citep{hunter:2007}, \textsc{numpy} \citep{harris_etal:2020}, \textsc{pandas} \citep{mckinney:2010}, \textsc{scipy} \citep{virtanen_etal:2020} \textsc{python} packages. 
\end{acknowledgements}

\
\bibliographystyle{aa}
\bibliography{cyg}


\begin{appendix}

\onecolumn

\section{Old Cygnus\texorpdfstring{\,}{}A model components}\label{app:cyga-old-model}
In Table~\ref{tab:old-cyga_comp_list} we report the components of the old Cyg\,A model used by the current LOFAR-EoR processing pipeline. The model catalogue is available at \url{https://raw.githubusercontent.com/lofar-astron/prefactor/master/skymodels/Ateam_LBA_CC.skymodel}.

\begin{table*}[h!]
\caption{The components of the old Cygnus\,A model obtained from VLA observations.}
\label{tab:old-cyga_comp_list}
\centering
\begin{tabular}{lcccccccc}
\toprule
Name & Type & RA & Dec & $I$ & $\alpha$ & MajorAxis & MinorAxis & Orientation \\
 & & & & (kJy) & & (arcsec) & (arcsec) & (deg) \\
\midrule
CygA\_4\_1GG & GAUSSIAN & $19^\text{h}59^\text{m}30\rlap{.}^\text{s}691$ & $+40^\circ43'54\rlap{.}''226$ & $4.263$ & $-0.8$ & 18.21 & 11.69 & 69.14 \\
CygA\_4\_2GG & GAUSSIAN & $19^\text{h}59^\text{m}22\rlap{.}^\text{s}263$ & $+40^\circ44'27\rlap{.}''859$ & $4.164$ & $-0.8$ & 20.15 & 87.32 & 103.82 \\
CygA\_4\_0GG & GAUSSIAN & $19^\text{h}59^\text{m}27\rlap{.}^\text{s}371$ & $+40^\circ44'06\rlap{.}''504$ & $1.768$ & $-0.8$ & 23.13 & 69.14 & 78.24 \\
CygA\_4\_5GG & GAUSSIAN & $19^\text{h}59^\text{m}28\rlap{.}^\text{s}833$ & $+40^\circ44'02\rlap{.}''168$ & $1.394$ & $-0.8$ & 23.60 & 0 & 36.24 \\
CygA\_4\_7GG & POINT & $19^\text{h}59^\text{m}30\rlap{.}^\text{s}024$ & $+40^\circ43'58\rlap{.}''388$ & $1.309$ & $-0.8$ & -- & -- & -- \\
CygA\_4\_3GG & GAUSSIAN & $19^\text{h}59^\text{m}24\rlap{.}^\text{s}073$ & $+40^\circ44'18\rlap{.}''226$ & $1.226$ & $-0.8$ & 25.73 & 0 & 35.31 \\
CygA\_4\_4GG & GAUSSIAN & $19^\text{h}59^\text{m}24\rlap{.}^\text{s}980$ & $+40^\circ44'09\rlap{.}''455$ & $0.865$ & $-0.8$ & 34.47 & 0 & 158.91 \\
CygA\_4\_6GG & GAUSSIAN & $19^\text{h}59^\text{m}25\rlap{.}^\text{s}564$ & $+40^\circ44'21\rlap{.}''868$ & $0.700$ & $-0.8$ & 21.55 & 0 & 66.46 \\
CygA\_4\_11GG & POINT & $19^\text{h}59^\text{m}28\rlap{.}^\text{s}553$ & $+40^\circ43'43\rlap{.}''492$ & $0.066$ & $-0.8$ & -- & -- & -- \\
CygA\_4\_8GG & POINT & $19^\text{h}59^\text{m}20\rlap{.}^\text{s}591$ & $+40^\circ44'17\rlap{.}''513$ & $0.057$ & $-0.8$ & -- & -- & -- \\
CygA\_4\_9GG & GAUSSIAN & $19^\text{h}59^\text{m}23\rlap{.}^\text{s}597$ & $+40^\circ44'44\rlap{.}''562$ & $0.051$ & $-0.8$ & 31.08 & 0 & 112.37 \\
CygA\_4\_10GG & POINT & $19^\text{h}59^\text{m}27\rlap{.}^\text{s}150$ & $+40^\circ43'49\rlap{.}''356$ & $0.001$ & $-0.8$ & -- & -- & -- \\
\bottomrule
\end{tabular}
\tablefoot{The flux density $I$ and the spectral index $\alpha$ are estimated at a reference frequency of 73.8\,MHz.}
\end{table*}

\section{Processing of the Cygnus\texorpdfstring{\,}{}A data set}\label{app:cyga-processing}

The initial data set consisted of 399 and 203 SBs for HBA-low and HBA-high, respectively, with frequency and time resolutions of 48.8\,kHz (i.e.\ 4 channels per SB) and 2\,s. In the following, we will briefly describe the flagging, averaging, and calibration that have been previously performed using \textsc{casa} \citep[Common Astronomy Software Applications;][]{mcmullin_etal:2007}. 

Firstly, the data set has been averaged to a single channel per SB (i.e.\ 195.3\,kHz of frequency resolution) and to an integration time of 2\,s. The SB at the edges of both bands are flagged because of strong RFI. Then, the calibration of the HBA-low data has been performed using the best model from \citet{mckean_etal:2016}, solving both phase and amplitude for each channel (i.e.\ one solution every 195.3\,kHz). The calibrated data has been imaged with the \textsc{casa} multiscale multifrequency synthesis deconvolution, decomposing the spectral variation with two terms of the logarithmic polynomial function (see Eq.~\ref{eq:logpol}). This model is used to calibrate the HBA-high data, which are then concatenated to the HBA-low data. 

The full data set has been self-calibrated with a two Taylor term model obtained from the combined bands and finally flagged, resulting in a total of 412 SBs spanning 111--181 and 210--249\,MHz.

\FloatBarrier 
\clearpage

\section{Simple NCP sky model}\label{app:simple-NCP_model}
In Table~\ref{tab:sim-model}, we report the central position and total flux density of the each of the 14 clusters that constitute the simple NCP sky model. We used this model for the simulated data set.

\begin{table*}[h!]
\caption{Details of the NCP model used in the simulation step.}
\label{tab:sim-model}
\centering
\begin{tabular}{lcccc}
\toprule
Cluster & RA & Dec & Flux density\\
& & & (Jy) \\
\midrule 
1 & $02^\text{h}59^\text{m}32\rlap{.}^\text{s}94$ & $+86^\circ25'28\rlap{.}''75$ & 40.66 \\
2 & $12^\text{h}57^\text{m}54\rlap{.}^\text{s}30$ & $+86^\circ38'13\rlap{.}''36$ & 33.73 \\
3 & $19^\text{h}06^\text{m}55\rlap{.}^\text{s}94$ & $+86^\circ42'41\rlap{.}''86$ & 30.11 \\
4 & $16^\text{h}23^\text{m}41\rlap{.}^\text{s}99$ & $+86^\circ30'09\rlap{.}''31$ & 28.94 \\
5 & $07^\text{h}32^\text{m}55\rlap{.}^\text{s}82$ & $+86^\circ36'33\rlap{.}''91$ & 27.96 \\
6 & $10^\text{h}00^\text{m}37\rlap{.}^\text{s}52$ & $+86^\circ23'21\rlap{.}''35$ & 27.40 \\
7 & $01^\text{h}49^\text{m}35\rlap{.}^\text{s}26$ & $+88^\circ25'47\rlap{.}''16$ & 26.49 \\
8 & $21^\text{h}17^\text{m}24\rlap{.}^\text{s}94$ & $+86^\circ20'27\rlap{.}''92$ & 25.32 \\
9 & $14^\text{h}35^\text{m}08\rlap{.}^\text{s}05$ & $+88^\circ22'37\rlap{.}''94$ & 25.06 \\
10 & $20^\text{h}00^\text{m}36\rlap{.}^\text{s}09$ & $+88^\circ47'30\rlap{.}''65$ & 23.62 \\
11 & $05^\text{h}00^\text{m}54\rlap{.}^\text{s}66$ & $+87^\circ05'35\rlap{.}''73$ & 19.10 \\ 
12 & $08^\text{h}13^\text{m}21\rlap{.}^\text{s}06$ & $+88^\circ40'41\rlap{.}''76$ & 18.82 \\
13 & $23^\text{h}06^\text{m}49\rlap{.}^\text{s}67$ & $+87^\circ04'30\rlap{.}''64$ & 9.19 \\
14 & $00^\text{h}44^\text{m}22\rlap{.}^\text{s}68$ & $+86^\circ22'36\rlap{.}''81$ & 6.55 \\
\bottomrule
\end{tabular}
\end{table*}

\section{Final power spectrum differences}\label{app:results-diff}

In Table~\ref{tab:merged-diff-results}, we report the differences $D_{ij}=\Delta_i^2 -\Delta_j^2$ and the associated $1\sigma$ uncertainties evaluated using Eq.~\eqref{eq:diff_ps3d}. Both the results from simulations (Sect.~\ref{sec:results-sim-sps}) and real observations (Sect.~\ref{sec:results-obs-sps}) are shown. In the last two rows, we also report the inverse-variance weighted mean of the differences $D_{ij}$ over the $k$-bins ranges and the associated $1\sigma$ uncertainties, evaluated as
\begin{equation}\label{eq:inv-var-wmean-diff}
\langle D_{ij}\rangle = \frac{\sum_k{\left(D_{ij}(k) \cdot D_{ij,\text{err}}^{-2}(k)\right)}}{\sum_k{D_{ij,\text{err}}^{-2}(k)}}\, ,\ \langle D_{ij} \rangle_{\text{err}} = \sqrt{\frac{1}{\sum_k{D_{ij,\text{err}}^{-2}}}}\,.
\end{equation}

\begin{table*}[h!]
\caption{Differences of spherical power spectrum pairs for simulated and observed NCP data, with the associated $1\sigma$ error.}
\label{tab:merged-diff-results}
\centering
\resizebox{1\textwidth}{!}{
\begin{tabular}{ccccccc}
\toprule
 & \multicolumn{3}{c}{Simulations} & \multicolumn{3}{c}{Observations} \\
\cmidrule(lr){2-4}\cmidrule(lr){5-7}
k & $\Delta_\text{old}^2-\Delta_\text{new}^2$ & $\Delta_\text{point}^2-\Delta_\text{new}^2$ & $\Delta_\text{point}^2-\Delta_\text{old}^2$ & $\Delta_\text{old}^2-\Delta_\text{new}^2$ & $\Delta_\text{point}^2-\Delta_\text{new}^2$ & $\Delta_\text{point}^2-\Delta_\text{old}^2$\\
($h\,\text{cMpc}^{-1}$) & (mK$^2$) & (mK$^2$) & (mK$^2$) & (mK$^2$) & (mK$^2$) & (mK$^2$) \\
\midrule
0.08        & $ +(7.10)^2 \pm (41.99)^2 $       & $ +(26.02)^2 \pm (42.94)^2 $      & $ +(25.03)^2 \pm (43.01)^2 $     & $ -(223.23)^2 \pm (2113.76)^2 $   & $ +(535.38)^2 \pm (2123.78)^2 $   & $ +(580.06)^2 \pm (2122.31)^2 $   \\ 
0.11        & $ -(5.17)^2 \pm (57.06)^2 $       & $ +(23.31)^2 \pm (57.48)^2 $      & $ +(23.88)^2 \pm (57.46)^2 $     & $ +(480.41)^2 \pm (1390.86)^2 $   & $ +(180.77)^2 \pm (1384.98)^2 $   & $ -(445.10)^2 \pm (1391.81)^2 $   \\ 
0.14        & $ +(30.91)^2 \pm (79.87)^2 $      & $ +(67.85)^2 \pm (81.24)^2 $      & $ +(60.40)^2 \pm (81.57)^2 $     & $ +(390.49)^2 \pm (957.66)^2 $    & $ +(331.09)^2 \pm (956.53)^2 $    & $ -(207.02)^2 \pm (960.52)^2 $    \\ 
0.19        & $ +(69.71)^2 \pm (115.49)^2 $     & $ +(239.41)^2 \pm (124.51)^2 $    & $ +(229.03)^2 \pm (125.15)^2 $   & $ -(189.67)^2 \pm (954.30)^2 $    & $ +(256.63)^2 \pm (956.06)^2 $    & $ +(319.12)^2 \pm (955.44)^2 $    \\ 
0.26        & $ +(82.28)^2 \pm (122.58)^2 $     & $ +(424.47)^2 \pm (142.62)^2 $    & $ +(416.42)^2 \pm (143.09)^2 $   & $ +(225.44)^2 \pm (661.87)^2 $    & $ +(248.76)^2 \pm (662.08)^2 $    & $ +(105.15)^2 \pm (663.02)^2 $    \\ 
0.35        & $ +(198.16)^2 \pm (175.76)^2 $    & $ +(861.36)^2 \pm (230.58)^2 $    & $ +(838.25)^2 \pm (231.88)^2 $   & $ +(203.28)^2 \pm (346.58)^2 $    & $ +(150.94)^2 \pm (345.95)^2 $    & $ -(136.16)^2 \pm (347.35)^2 $    \\ 
0.46        & $ +(55.75)^2 \pm (242.08)^2 $     & $ +(185.30)^2 \pm (243.67)^2 $    & $ +(176.71)^2 \pm (243.82)^2 $   & $ +(179.24)^2 \pm (539.60)^2 $    & $ +(835.65)^2 \pm (553.24)^2 $    & $ +(816.20)^2 \pm (553.84)^2 $    \\ 
0.63        & $ -(24.35)^2 \pm (349.32)^2 $     & $ +(543.64)^2 \pm (358.42)^2 $    & $ +(544.19)^2 \pm (358.40)^2 $   & $ +(1188.90)^2 \pm (887.23)^2 $   & $ +(2887.20)^2 \pm (964.72)^2 $   & $ +(2631.06)^2 \pm (976.20)^2 $   \\ 
0.85        & $ +(61.80)^2 \pm (495.40)^2 $     & $ +(619.05)^2 \pm (502.52)^2 $    & $ +(615.96)^2 \pm (502.59)^2 $   & $ +(1654.94)^2 \pm (1074.88)^2 $  & $ +(4666.05)^2 \pm (1228.38)^2 $  & $ +(4362.70)^2 \pm (1242.11)^2 $  \\ 
1.15        & $ -(41.87)^2 \pm (718.76)^2 $     & $ +(118.13)^2 \pm (718.93)^2 $    & $ +(125.33)^2 \pm (718.91)^2 $   & $ +(982.77)^2 \pm (1014.69)^2 $   & $ -(923.45)^2 \pm (1001.99)^2 $   & $ -(1348.55)^2 \pm (1008.87)^2 $  \\ 
\midrule
0.08--0.63   & $ +(17.38)^2 \pm (38.57)^2 $      & $ +(56.30)^2 \pm (39.38)^2 $      & $ +(54.20)^2 \pm (39.43)^2 $     & $ +(256.98)^2 \pm (324.09)^2 $    & $ +(463.29)^2 \pm (325.06)^2 $    & $ +(391.35)^2 \pm (326.23)^2 $    \\ 
0.63--1.15   & $ +(13.94)^2 \pm (326.95)^2 $     & $ +(547.32)^2 \pm (334.35)^2 $    & $ +(547.11)^2 \pm (334.35)^2 $   & $ +(1259.39)^2 \pm (741.59)^2 $   & $ +(2663.22)^2 \pm (788.59)^2 $   & $ +(2361.20)^2 \pm (796.34)^2 $   \\ 

\bottomrule
\end{tabular}}
\tablefoot{Differences were estimated using Eq.~\eqref{eq:diff_ps3d} at given $k$-bins. In the last two rows, we report the inverse-variance weighted mean of the differences over the indicated $k$-bins ranges, with the associated standard error, estimated using Eq.~\eqref{eq:inv-var-wmean-diff}.}
\end{table*}

\end{appendix}

\end{document}